\documentclass[amsthm]{elsart}

\usepackage{yjsco}
\usepackage{natbib}

\usepackage{multirow}
\usepackage{mathrsfs}
\usepackage{graphicx}  %% note spelling
\usepackage{float}
\usepackage{amssymb}
\usepackage{amsmath}
\usepackage{cases}
\usepackage{amsfonts}
\usepackage{mathrsfs}

\def\qed{\hfil {\vrule height5pt width2pt depth2pt}}

\def\qed{\hfil {\vrule height5pt width2pt depth2pt}}
\def\bref#1{(\ref{#1})}

\def\Z{{\mathbb{ Z}}}
\def\R{{\mathbb{R}}}
\def\Q{{\mathbb{Q}}}
\def\V{{\mathbb{V}}}
\def\IQ{{\mathbb{I}\mathbb{Q}}}

\def\res{{\hbox{\rm{Res}}}}

\def\C{{\mathbb{C}}}

\def\I{{\mathbb{I}}}

\def\sep{{\rm{sep}}}

\def\O{{\tilde{\mathcal{O}}}}
\def\OB{{\tilde{\mathcal{O}}_B}}

\def\LL{{\mathcal L}}

\begin{document}
\begin{frontmatter}

\title{A Generic Position Based Method for Real Root Isolation of Zero-Dimensional Polynomial Systems }
%\footnote{The work is partially supported by NKBRPC (2011CB302400), NSFC Grants (11001258, 60821002), and
%China-France cooperation project EXACTA (60911130369) and ECCA. \hspace{50mm}
%Email addresses:  jcheng@amss.ac.cn (J.S. Cheng), jinkai@amss.ac.cn (K. Jin)}}
\footnotetext{%The corresponding author: Jin-San Cheng, Fax: +86-10-6263-0706.\\
Email:jcheng@amss.ac.cn(Jin-San Cheng), jinkaijl@163.com(Kai Jin). }

%\numberofauthors{1}
 %  in this sample file, there are a *total*
\author{Jin-San Cheng,  Kai Jin}

\address{   KLMM, Institute of Systems Science, AMSS, Chinese Academy of Sciences }

%\date{}

%\maketitle
\begin{abstract}
We improve the local generic position method for isolating the real
roots of a zero-dimensional bivariate polynomial system with two polynomials and extend the method to general zero-dimensional polynomial systems. The method mainly
involves resultant computation and real root isolation of univariate
polynomial equations. The roots of the system have a linear
univariate representation. The complexity of the method is
$\tilde{O}_B(N^{10})$ for the bivariate case, where $N=\max(d,\tau)$, $d$ resp., $\tau$ is an upper bound on the
degree, resp., the maximal coefficient bitsize of the input polynomials. The algorithm is certified with probability 1 in the multivariate case. The
implementation shows that the method is efficient, especially for
bivariate polynomial systems.
\end{abstract}

%
%
%% A category with the (minimum) three required fields
%\category{I.1.2}{SYMBOLIC AND ALGEBRAIC
%MANIPULATION}{Algorithms}[Algebraic algorithms ]
%%A category including the fourth, optional field follows...
%%\category{D.2.8}{Software Engineering}{Metrics}[complexity measures, performance measures]
%
%\terms{Algorithm}
%
%\keywords{Polynomial system,
%real root isolation, symbolic-geometric, generic position.} % NOT required for Proceedings
%
%
\end{frontmatter}
\section{Introduction}

Real root isolation of zero-dimensional polynomial systems is a fundamental problem in
symbolic computation and it has many applications.
The problem has been studied for a long time and there are a lot of
results.  One can compute the real roots of a zero-dimensional polynomial system by symbolic methods, numeric methods and symbolic-numeric methods. In context of symbolic methods, we can mention the characteristic set methods, Gr\"obner basis
methods, the resultant methods and so on. In this paper, we focus on the resultant methods. We consider the zero-dimensional system as $\{f_1,\ldots,f_m\}\subset \Z[x_1,\ldots,x_n]$, where $\Z$ is the ring of integers.

The idea of this paper comes from a geometric property of the roots of a polynomial system: generic position. Generic position was used in the polynomial system solving for a long time
(\cite{alonso1,bw93,canny1,chenglgp,eq2-dio,giu,resol,gh91,koba1,rouillier-rur,tan,yoko1}). Let's explain it for the bivariate case.
Simply speaking, a zero-dimensional bivariate system is said to be in a {\bf generic
position} if we can find a complex plane, say the $x$-axis, such that
different complex zeros of the system are projected to different complex points on
the complex $x$-axis.
%A {\bf local generic position} method is proposed to
%isolate the real roots of a zero-dimensional bivariate polynomial system in \cite{chenglgp}.
In the rest of this paper, when we say root(s), we mean real root(s) if there is no special illustration.

%If $\Sigma$ is in a generic position,  the roots of an
%equation system $\Sigma=0$ have a rational univariate representation
%(RUR) \cite{rouillier-rur}:
%%
%\begin{equation}\label{eq-rur} t(u)=0, x=R_1(u), y = R_2(u)\end{equation}
%%
%where $u$  is a new parameter,  $t(u)\in\Q[u]$ and $R_1(u),R_2(u)$
%are rational functions. As a consequence, solving multi-variate
%equations is reduced to solving a univariate equation $t(u)=0$ and
%to substituting the roots of $t(u)=0$ into rational functions.
%%
%This approach still has the following problem: for an isolation
%interval $[a,b]$ of a real root $\alpha$ of $t(u)=0$, to determine
%the isolation interval of $R_1(\alpha)$ and $R_2(\alpha)$ under a
%given precision is not a trivial task.
%%
%%{\bf In fact, to get the root is not difficult, but the gays didn't
%%use the right method. If they use sleeve method, it is convenient to
%%get the isolating interval. And Fabrice just used his current method
%%in his program, not in his paper, or I didn't see the paper. So I
%%suggest to delete the last paragraph.}
%%%
%%But this method is only fit for the systems in generic position,
%%that is, the method is not a complete method.
%The local generic position method proposed in this paper will remedy
%this drawback.

Solving bivariate polynomial systems is widely studied in recent
years (\cite{buse,chenglgp,cor,eri,emi,eq2-dio,emeli1,hong2,qin}).
Most of these methods projected the systems to two directions
($x$-axis, $y$-axis) and identified whether a root pair (one
$x$-coordinate and one $y$-coordinate) was a true root or not
(\cite{eq2-dio,emeli1,hong2,qin}). In (\cite{buse,cor}), they projected
the roots of the bivariate system to $x$-axis, using a matrix
formulation, and lifted them up to recover the roots of the original
system. The multiplicity of the roots are also considered.

A {\bf
local generic position} method was proposed to isolate the real roots
of a zero-dimensional bivariate polynomial system in
(\cite{chenglgp}). In the local generic position method, the roots of
a zero-dimensional bivariate polynomial system
$\Sigma=\{f(x,y),g(x,y)\}$ are represented as linear combinations of
the roots of two univariate polynomial equations
$R_1(x)=\res_y(f,g)=0$ and
$R_2(x)=\res_y(f(x+s\,y,y),g(x+s\,y,y))=0$: {\small
$$\{x=\alpha,\,y=\frac{\beta-\alpha}{s}\,|\,\alpha\in
\V(R_1(x)),\beta\in \V(R_2(x)),|\beta-\alpha|<S\},
$$}
where $s, S$ are constants satisfying certain given conditions. Each
root $(\alpha,\beta)$ of $\Sigma=0$ is projected in $R_2(x)=0$ such
that the corresponding root is in a neighborhood of $\alpha: E=\{v|
|v-\alpha|<S\}$. All the roots of $R_2(x)=0$ in $E$ correspond to
the roots of $\Sigma=0$ on the fiber $x=\alpha$. Thus we can recover
the $y$-coordinates of the roots of $\Sigma=0$ from the roots of
$R_2(x)=0$.  The multiplicities of the roots of $\Sigma=0$ are also
preserved in the corresponding roots of $R_2(x)=0$. The
implementation of the method showed that it is efficient and stable
when compared to the best methods at that time, especially when the
system had multiple roots. The local generic position method has a bottleneck. When some of
the roots of $R_1(x)$ are very close, $s$ will be very small. Thus
computing $R_2(x)$ and isolating its roots is time-consuming. Sometimes, it is more than $90\%$ of the total computing time! The rate increases when the
degrees of the polynomials in the systems increase.

The contribution of the paper is that we present a method to overcome
the bottleneck of the local generic position method and extend the
method to general zero-dimensional multivariate polynomial system
mainly involving resultant computation and univariate polynomial
root isolation, which is easy to implement. We also analyze the
complexity of the algorithm for the bivariate case. We compare our implementation with several other efficient
related softwares, such as local generic position method(\cite{chenglgp}), Hybird method(\cite{hong2}),
Discovery(\cite{dis}) and Isolate (\cite{rouillier-rur}). The results show that our algorithm is
efficient, especially in bivariate case.

In order to overcome the
drawback of the local generic position method, we present a method to search for a better $s$ with a small
bitsize and present another way to recover the roots of the system. This is
the main contribution of the paper. Finding the correspondence
between the roots of $\Sigma=0$ and $R_2(x)=0$, we can recover the
roots of $\Sigma=0$. It works as follows. First, we compute $R_1(x)$
and its roots. From the isolating intervals of the roots of
$R_1(x)=0$, we get the root isolating interval candidates of $f=g=0$
by computing the roots of interval polynomials. We compute a rational number $s$ such that any two isolating interval candidates
are not overlapping under a linear transformation $\varphi: (x,y)\rightarrow (x+s\,y,y)$ and
$\{\varphi(f),\,\varphi(g)\}$ is in a generic position. Then for each isolating interval candidate
$K=[a,b]\times [c,d]$, we can isolate the roots of $R_2(x)=0$ in the
interval $\pi_y(\varphi(K))$ ($\pi_y: (x,y)\rightarrow (x)$) to recover the isolating intervals of
$f=g=0$. The multiplicity(ies) of the root(s) of the system in $K$ is(are)
the multiplicity(ies) of the corresponding root(s) in $\pi_y(\varphi(K))$. The bivariate polynomial system with several polynomials can be solved using the method with a little modification (see Section 4).
%We modify the method to solve bivariate systems with more than 2 polynomials.

We extend the method to zero-dimensional polynomial systems in
the multivariate case. Let's consider the trivariate  case for an
example. For a zero-dimensional polynomial system
$\{f_1,f_2,f_3\}\subset \Z[x,y,z]$, we can get a bivariate
polynomial system $\{g_1,g_2\}\subset \Z[x,y]$, where
$g_1=\res_z(f_1,f_2), g_2=\res_z(f_1,f_3)$. Isolating the roots of
$\{g_1,g_2\}$, using the isolating intervals to construct interval
polynomials for $f_1,f_2,f_3$,  isolating the roots of these
interval polynomials, we can get the root isolating interval
candidates of the system $\{f_1,f_2,f_3\}$. For all the root
isolating interval candidates, we separate them into different groups such that the
first coordinates of the isolating boxes in each group are the same. We compute an $s$ such that for each
group, the last two coordinates of the corresponding roots of
$\{f_1(x,y+s\,z,z),f_2(x,y+s\,z,z),f_3(x,y+s\,z,z)\}$ in the group
are in a generic position. Solving
$\mathcal{P}=\{\res_z(f_1(x,y+s\,z,z),f_2(x,y+s\,z,z)),\res_z(f_1(x,y+s\,z,z),f_3(x,y+s\,z,z))\}$,
we can check whether the root candidates of $\{f_1,f_2,f_3\}=0$
containing its real roots or not from the roots of $\mathcal{P}=0$.
Sometimes we need to take a linear combination of $f_i$'s to
construct a new system to ensure that the two projection polynomials
form a zero-dimensional system. In a similar way, we can solve a
general zero-dimensional polynomial system. This method usually works
well for the systems with 2 or 3 variables.
%The reason is that the method is of double exponential complexity for the multivariate case.
%Of course, we require the number of variables and the number of equations are the same.

The complexity of the bivariate system
solving is studied before. One is $\tilde{O}_B(N^{12})$ (\cite{eq2-dio}), the other is
$\tilde{O}_B(N^{8})$ (\cite{emeli}). Ours is $\tilde{O}_B(N^{10})$, where $N$ is
the maximum between the degree bound and the bitsize bound of the
coefficients of the polynomials in the system.

The rest of this paper is organized as follows. In Sections 2 and 3,
the basic tools related to interval polynomials and generic position
are introduced. In Section 4, we present the improved bivariate
systems solving method. In Section 5, the improved method is
extended to general 0-dimensional system. In Section 6, we give the
complexity analysis of this algorithm. Experimental results are
presented in Section 7.

\section{Interval polynomial and its real roots}
\label{sec-1} In this section, we will show how to construct an
interval polynomial related to a polynomial and how to compute the
real roots of an interval polynomial. Interval methods were also used to solve polynomial systems before (\cite{man,stahl,mour}).

Let $\Q,\R,\C$ be the fields of rational numbers, real numbers and
complex numbers respectively.

Denote $\V(f)$ as the zeros in $\C^n$ of $f\in \Q[x_1,\ldots,x_n]$
and $\V_\R(f)=\V(f)\cap \R^n$. Here $f$ can also be a polynomial
system.

Given  $f=a_0+a_1 x+...+a_n x^n\in \Z[x]$, we can rewrite it in Horner form.
$$f_h=a_0+(a_1+(a_2+\ldots+(a_{n-1}+a_n x)\cdots x)x)x.$$
If $a_i\in \Z[x_1]$ and we rewrite it in Horner form, then $f_h\in \Z[x_1,x]$ is a bivariate polynomial in Horner form with order $x_1\prec x$.
Recursively, we can rewrite a multivariate polynomial $f\in\Z[x_1,\ldots,x_n]$ in Horner form in a fixed variable order $x_1\prec x_2\prec\cdots\prec x_n$.

Let $f\in \Z[x_1,\ldots,x_n,x]$ and rewrite it as below
$$f=h_0+h_1 x+\ldots+h_m x^m,$$
where $h_i\in\Z[x_1,\ldots,x_n](i=0,\ldots,m)$ are in Horner form in
a fixed variable order $x_1\prec x_2\prec\cdots\prec x_n$.

Let $\IQ$ denote the set of intervals whose endpoints are rational
numbers and $\IQ^n$ denote a set of intervals as
$I_1\times\cdots\times I_n$, where $I_i\in\IQ$. Let
$\I=I_1\times\cdots\times I_n\in \IQ^n$. Evaluating $\I$ for
$x_1,\ldots,x_n$ in $h_i (i=0,\ldots,m)$, we can derive an interval,
say $A_i=h_i(\I)=[a_i,b_i]$. One can find more details on the properties and techniques of interval
arithmetics in (\cite{mor,stahl}). It is clear that $h_i(x_0)\in
h_i(\I)=A_i$. $h_i(x_0)$ is strictly inside $(a_i,b_i)$ if not
all $a_i=b_i$ for $i=0,\ldots,m$. We can derive an interval
polynomial for $f$ related to $\I$.
$$f(\I,x)=\sum_{i=0}^m A_i x^i=\sum_{i=0}^m [a_i,b_i] x^i.$$

Consider $\V_\R(f(x_1,\ldots,x_n,x))$ in the region $\I\times
[0,+\infty]$. Note that we can get the related information of $\V_\R(f(x_1,\ldots, x_n,x))$ in
the region $\I\times [-\infty,0]$ by considering
$f(x_1,\dots,x_n,-x)=0$ in the region $\I\times [0,+\infty]$. Denote
\begin{eqnarray}
f_\I^u(x)=b_0+b_1 x+\ldots+b_m x^m,
f_\I^d(x)=a_0+a_1 x+\ldots+a_m x^m. \nonumber
\end{eqnarray}
We can find that $f_\I^u(x), f_\I^d(x)$ are the bounding polynomial of the interval polynomial $f(\I,x)$, that is, the region defined by $f(\I,x)=0$ are bounded by $f_\I^u(x)=0, f_\I^d(x)=0$.

The following inequality holds (see \cite{chengjsc}).
\begin{equation}\label{eq-mono}
\frac{\partial^k  f_\I^d(x)}{\partial x^k}< \frac{\partial^k f(x_0,x)}{\partial x^k}< \frac{\partial^k f_\I^u(x)}{\partial x^k}, \forall x\ge 0, \forall x_0\in \I, k=0,1.
\end{equation}

\begin{defn}
 We call an open interval $(s,t)$ {\bf a real root} of $f_\I(x)=0$ if
\begin{enumerate}
\item  $s,t (s<t)$ are real root(s) of $f_\I^u(x)f_\I^d(x)=0$ or $0, +\infty$;
\item  ${\rm sign}(f_\I^u(x)) {\rm sign}(f_\I^d(x))<0, \,\,\forall x\in (s,t)$.
\end{enumerate}
\end{defn}
\begin{lem}
Use the same notations as above. Any real root of $f(x_0,x)=0$ are
inside some real root of $f_\I(x)=0$ for $x_0\in\I$.
\end{lem}
\noindent{\bf Proof.} Let $\bar{x}\ge0$ be a real root of $f(x_0,x)=0$. By
\bref{eq-mono}, $f_\I^d(\bar{x})< f(x_0,\bar{x})=0<
f_\I^u(\bar{x})$. Thus $\bar{x}$ is in some real root of
$f_\I(x)=0$. \qed

The lemma shows that all the real roots of $f(x_0,x)=0$ are
contained in the real roots of $f_\I(x)=0$.

\begin{defn}[\cite{chengjsc}]
We call $f_\I(x)$ {\bf monotonous} in its real root $(s,t)$ if
$$\begin{cases}
s\in \V_\R(f_\I^u), t\in \V_\R(f_\I^d), \hbox{ and } \V_\R(\frac{\partial f_\I^d}{\partial x})\cap (s,t)=\emptyset,\,\,\,  (*)\\
\hbox{ or }\\
s\in \V_\R(f_\I^d), t\in \V_\R(f_\I^u), \hbox{ and } \V_\R(\frac{\partial f_\I^u}{\partial x})\cap (s,t)=\emptyset.\,\,\,  (**)
\end{cases}
$$
\end{defn}
Note that $(*)$ means the bounding polynomial $f^d_\I(x)$ is strictly increasing in $(s,t)$ and $(**)$ means the bounding polynomial $f^u_\I(x)$ is strictly decreasing in $(s,t)$.

\begin{lem}\label{lem-mono}
If $f(\I,x)$ is {\bf monotonous} in $(s,t)$, then $f(x_0,x)=0$ has exactly one real root in $(s,t)$ for any $x_0\in \I$.
\end{lem}
\noindent{\bf Proof.} At first, we prove that there exists one real root.
Assume that $(*)$ holds, the proof for $(**)$ is similar. For any
$x_0\in \I$, since \bref{eq-mono} holds, $f(x_0,s)<f_\I^u(s)=0$ and
$f(x_0,t)>f_\I^d(t)=0$. Thus $f(x_0,x)=0$ has real roots in $(s,t)$.
We will prove that there is only one real root. Since
$V(\frac{\partial f_\I^d}{\partial x})\cap (s,t)=\emptyset$,
$f_\I^d(x)$ is monotonous in $(s,t)$. From \bref{eq-mono}, we know
$f(x_0,x)$ is also monotonous in $(s,t)$ (see the detailed proof in
\cite{chengjsc}). Thus it has only one real root in $(s,t)$.\qed

Now we construct an effective version for the real roots of $f_\I(x)=0$. We will use rational numbers $a,b$ to replace algebraic numbers $s,t$ such that $(s,t)\subset[a,b]$. We will show how to construct the effective roots in $[0,\,\infty)$ with the following algorithm.
\begin{alg} \label{alg-candi} Compute the effective real roots of $f_\I(x)=0$. Input: $f_\I(x)$. Output: the effective real roots of $f_\I(x)=0$.
\end{alg}
\begin{enumerate}
\item Isolate the real roots of
$f_\I^d(x)$ and $f_\I^u(x)$, denoted by $\mathcal
{I}^d=\big\{I^d_i=[a^d_i,\,b^d_i]|i=1,\ldots,m_1\big\}$ and
$\mathcal {I}^u=\big\{I^u_i=[a^u_i,\,b^u_i]|i=1,\ldots,m_2\big\}$
respectively. Assume $\mathcal {I}^d\bigcup\mathcal
{I}^u=\big\{[\bar{a}_i,\bar{b}_i]|i=1,\ldots,m\big\}$, where $0\leq
\bar{a}_1\leq\bar{b}_1<\cdots<\bar{a}_i\leq\bar{b}_i<\cdots<\bar{a}_m\leq\bar{b}_m$.

\item If $f_\I^d(0)f_\I^u(0)\leq 0$, add $[0,0]$ as the first element of $\mathcal {I}^d\bigcup\mathcal
{I}^u$ if it is not contained in and $f^d_I(\frac{\bar{a}_1}{2})f^u_I(\frac{\bar{a}_1}{2})\ge0$; set $\bar{a}_1:=0$ if
$f^d_I(\frac{\bar{a}_1}{2})f^u_I(\frac{\bar{a}_1}{2})<0$.

 \item Denote $J:=[\bar{a}_1,\infty)$. For $i$ from 1 to $m-1$, do\\
  Denote $c_i:=\frac{\bar{b}_i+\bar{a}_{i+1}}{2}$. If
  $f_\I^d(c_i)f_\I^u(c_i)>0$, then delete the open interval $(\bar{b}_i,\,\bar{a}_{i+1})$
  from $J$, that is $J:=J\setminus(\bar{b}_i,\,\bar{a}_{i+1})$.\\
Denote $c_m=b_m+1$. If $f_\I^d(c_m)f_\I^u(c_m)>0$, $J:=J\setminus(\bar{b}_i,\,\infty)$. Else, compute a bound on $x$, say $b$, $J:=J\setminus(b,\,\infty)$.
\item
  After this process, the obtained interval set $J\triangleq\{[\tilde{a}_i,\tilde{b}_i]|i=1,\ldots,m_0\}$ is the
  effective roots of $f_\I(x)=0$. Output $J$.
\end{enumerate}
The correctness and termination of the algorithm is clear. We would
like to mention that when $f_\I^d(c_m)f_\I^u(c_m)<0$ in Step 3, the
signs of the leading coefficients of $f_\I^d(x),f_\I^u(x)$ are
different. We can check that whether $x_0$ vanishes at the leading
coefficient of $f$ w.r.t. $x$ easily for the case $f$ is a bivariate
polynomial. Then we can remove the leading term of $f$ w.r.t. $x$
when we construct the interval polynomial for $\I$. In doing so, we can
ensure that $x_0$ does not vanish at the new polynomial related to
$f$. Thus we can ensure that the leading coefficients of
$f_\I^d(x),f_\I^u(x)$ have the same sign. Note that sometimes a
refinement of $\I$ may be necessary. In fact, a similar checking can
be done for the multivariate case though it is much complicated than
the bivariate case. But for all the case, we can compute a
univariate polynomial in $x$ by resultant computation to get its
largest positive root as the bound.

Let $\Sigma=\{f_1,\ldots,f_m\}$ be a zero-dimensional polynomial system. $\I=I_1\times\cdots\times I_{n-1}$ is an isolating interval for a real root $\alpha=(\alpha_1,\ldots,\alpha_{n-1})$ of an $(n-1)$  projection system of $\Sigma$ (see Section 5), where the leading coefficients of $f_i$'s in $x_n$ are not all vanishing on $\alpha$. Otherwise, a linear coordinate transformation on $\Sigma$ can avoid it. Let $J_1,\ldots, J_k$ be the intersection of the effective real roots of $f_i(\I,x_n)=0, i=1,\ldots,m$. Thus $J_i$ are bounded. We call $I_1\times\cdots\times I_{n-1}\times J_j, j=1,\ldots,k$ the {\bf real root candidates} of $\Sigma=0$ (w.r.t. $\alpha$).

\section{Generic position}
In this section, we will show how to compute an $s$ such that a shear mapping
$$\varphi_{s,n}: (x_1,\ldots,x_{n-2},x_{n-1},x_n)\rightarrow (x_1,\ldots,x_{n-2},x_{n-1}+s\,x_n,x_n).$$
on a zero-dimensional polynomial system is in a generic position w.r.t. $x_{n-1},x_n$ (See Definition 3.2).

At first, we will consider a bivariate polynomial system. Let $f,g\in \Z[x,y]$ such that $\gcd(f,g)=1$. We say the system $\{f,g\}$ is in a {\bf generic position} w.r.t. $y$ if\\
%\indent 1) $\gcd(f,g)=1$.\\
\indent 1) The leading coefficients of $f$ and $g$ w.r.t. $y$ have no common factors.\\
\indent 2) Let $h$ be the resultant of $f$ and $g$ w.r.t. $y$. For any $\alpha\in \C$ such that $h(\alpha)=0$, $f(\alpha,y),g(\alpha,y)$
 have only one common zero in $\C$.

 Since we isolate the real roots of the system, the condition $\alpha\in \C$ can be revised as $\alpha\in \R$.

 Let $\pi_i (1\le i<n)$ be the projection map:
\begin{equation}\label{eq-pi0}
\pi_i: (z_1,\ldots,z_n)\longrightarrow (z_1,\ldots,z_i).\end{equation}

For a polynomial system $\Sigma\subset \Z[x_1,\ldots,x_n]$, we denote
$$\pi_i(\Sigma)=\Sigma\cap\Z[x_1,\ldots,x_i],$$
that is, the polynomial set in the ideal generated by $\Sigma$ with
only the variables $x_1,\ldots,x_i$.

We denote
$\varphi_{s,2}(f(x,y))=f(x+s\,y,y)$ below for convenience.

 Let $J_i=[a_i,b_i]\times[c_i,d_i]\in\IQ^2,i=1,2$.
Taking the map on $J_i$, we have
$$\varphi_{s,2}(J_i)=
\begin{cases}
[a_i+s\,c_i,b_i+s\,d_i]\times[c_i,d_i], s\ge 0, \\ [a_i+s\,d_i,b_i+s\,c_i]\times[c_i,d_i],s<0.
\end{cases}
$$

We denote
\begin{equation}\label{eqn-pro}
\pi_1(\varphi_{s,2}(J_i))=
\begin{cases}
[a_i+s\,c_i,b_i+s\,d_i], s\ge 0, \\ [a_i+s\,d_i,b_i+s\,c_i],s<0.
\end{cases}
\end{equation}

We say an $s$ is {\bf generic} w.r.t. $J_1,J_2$ if
$\pi_1(\varphi_{s,2}(J_1))\cap \pi_1(\varphi_{s,2}(J_2))=\emptyset.$
We say an interval or an interval set $S\subset\R$ is {\bf generic}
w.r.t. $J_1,J_2$ if $\forall s\in S, \pi_1(\varphi_{s,2}(J_1))\cap
\pi_1(\varphi_{s,2}(J_2))=\emptyset.$

It is obvious that for any point $P_i\in J_i, i=1,2$,
$\varphi_{s,2}(P_1)$ and $\varphi_{s,2}(P_2)$ will not overlap if
$s$ is {\bf generic} w.r.t. $J_1,J_2$.

Let $\mathcal{J}$ be a list of finite boxes as $J_i$. We say an
interval set $S\subset\R$ is {\bf non-generic} w.r.t. $\mathcal{J}$
if $\forall s\in S$, $\exists$ two boxes $J_1,J_2\in \mathcal{J}$,
$\pi_1(\varphi_{s,2}(J_1))\cap
\pi_1(\varphi_{s,2}(J_2))\neq\emptyset$. We call also $S$ a
non-generic interval set w.r.t. $\mathcal{J}$.

In order to compute $S$, we need to compute a non-generic interval set related to $J_1,J_2$, which can be achieved by solving the inequalities related to
$\pi_1(\varphi_{s,2}(J_1))\cap \pi_1(\varphi_{s,2}(J_2))=\emptyset$.
We will show an example to illustrate it.

\begin{exmp}
We will show how to compute a non-generic interval set for two boxes
$J_i\in \IQ^2, i=1,2$, where $J_1=[1,2]\times[3,4], J_2=[5,6]\times
[10,11]$. When $s\ge0$,
$T_1=\pi_1(\varphi_{s,2}(J_1))=[1+3\,s,2+4\,s],
T_2=\pi_1(\varphi_{s,2}(J_2))=[5+10\,s,6+11\,s]$ and
$2+4\,s<5+10\,s$. Thus $T_1\cap T_2=\emptyset$. When $s<0$,
$T_1'=\pi_1(\varphi_{s,2}(J_1))=[1+4\,s,2+3\,s],
T_2'=\pi_1(\varphi_{s,2}(J_2))=[5+11\,s,6+10\,s]$. The conditions
that $T_1'\cap T_2'=\emptyset$ are $2+3\,s<5+11\,s$ or
$6+10\,s<1+4\,s$. Solving them, we have $-3/8<s<0$ or $s<-5/6$. Thus
the condition that $T_1'\cap T_2'\neq\emptyset$ is $-5/6\le s\le
-3/8$. So the generic interval set for $J_1,J_2$ is $[[-5/6,-3/8]]$. And the non-generic interval set is $[(-\infty,-5/6), (-3/8,+\infty)]$.

\end{exmp}

\begin{defn}
We say a zero-dimensional polynomial system $\Delta\subset\Z[z_1,\ldots,z_n,x,y]$ is in {\bf a generic position w.r.t. $x,y$ in order $x\prec y$} (generic position to $x,y$ for short) if
for any (complex) zero $P$ of $\pi_n(\Delta)$, all the (complex) zeros of the system $\Delta$ on $P$ have distinct $x$-coordinates.
\end{defn}
For the definition above, since we consider only real roots of the
system in this paper, we can revise the condition as $\forall
P\in\V_\R(\pi_n(\Delta))$, $(P,\alpha_1,\alpha_2)$ is a root of
$\Delta$ and $\alpha_1\in\R$,  there is only one common complex root
of $\Delta$ on the fiber $(x_1,\ldots,x_n,x)=(P,\alpha_1)$.

Let $\beta\in\V_\R(\pi_{n-2}(\Sigma))$ and $\I$ the isolating interval for $\beta$. $\gamma_i, i=1,...,k$ are all the real roots of $\pi_{n-1}(\Sigma)$ at $\beta$ and $\I\times J_i$ are the corresponding isolating intervals of $(\beta,\gamma_i)$. Let $\I\times J_i\times K_{i,j}$ be all the real root candidates of $\Sigma$ w.r.t. $\beta$, where $i=1,\ldots,k, j=1,\ldots,t_i$, $t_i (1\le i\le k)$ are positive integers. We can compute a non-generic interval set w.r.t. $\{J_i\times K_{i,j}\}$, denoted as $S_\beta$. We take the union of this kind of intervals for all possible $\beta\in\V_\R(\pi_{n-2}(\Sigma))$. We can get a non-generic interval set
\begin{equation} \label{eq-si}
S=\cup_{\beta\in\V_\R(\pi_{n-2}(\Sigma))} S_\beta.
\end{equation}
Since the root candidates are finite and bounded, $\R\setminus S\neq
\emptyset$ if the isolating boxes are not very big. We can refine
the isolating boxes if needed. Our aim is to choose an
$s\in\R\setminus S$ such that the bitsize of $s$ is as small as
possible. The reason is that when taking a shear mapping on $f_i
(i=1,\ldots,n)$, the bitsizes of the coefficients of
$\varphi_{s,n}(f_i)$ are expected to be as small as possible. Thus the time (or you can say, the bit complexity) of computing resultants and the roots of the univariate polynomial equations is shorter (smaller). A
possible way is that choose a rational number $s$ in $\R\setminus S$
such that its bitsize is as small as possible. That is,
\begin{equation} \label{eq-s}
0\neq s\in \Q\setminus S, \hbox{ and } \mathcal{L}(s)\le \mathcal{L}(t), \forall t\in \Q\setminus S,
\end{equation}
where $\mathcal {L}(a)$ is the maximal bitsize of the numerator and
the denominator of $a\in \Q$. Of course, choose the best $s$ as
(\ref{eq-s}) is not easy. We can choose one that looks good.
Usually, we can choose $s$ as below:
$$0\neq s\in \Z\setminus S, \hbox{ and } |s|\le |t|, \forall t\in \Z\setminus S.$$

We would like to mention that since $\{J_i\times K_{i,j}\}$ contain
all the real roots of $\Sigma$ at $\beta$, the real roots of
$\varphi_{s,n}(\Sigma)$ at $\beta$ do not overlap when projected to
$x_{n-1}$-axis. So the method presented here computes a generic
position with respect to all the real roots. But it is not a
guaranteed generic position for all the complex roots since we
compute only the real roots. Of course, we can compute a guaranteed
generic position by computing the isolating interval of all the
complex roots with the method in \cite{chenglur}. But the aim of this paper is
to find all the real roots of the given system efficiently. With the method
above, the roots of the system is probability 1 in a generic
position w.r.t. $x_{n-1},x_n$ in order $x_{n-1}\prec x_n$. The
reason is that there may exist a fiber
$(x_1,\ldots,x_{n-1})=(\beta,\gamma_i)$ such that
$f_j(\beta,\gamma_i,x_n)=0$ for $j=1,\ldots,n$ have common conjugate
complex roots. Thus when we do certification of the real root candidates, some empty candidates may be regarded as containing real roots. But most of this case can avoid when we compute the
root candidates by interval arithmetic.

The following lemma is obvious.
\begin{lem} Let $\Sigma\subset\Z[x_1,\ldots,x_n] (n\ge 2)$. If we compute an integer $s$ as above from its real root candidates, then $\varphi_{s,n}(\Sigma)$ is in a generic position w.r.t. $x_{n-1},x_n$ in order $x_{n-1}\prec x_n$ with probability 1, where
$$\varphi_{s,n}:=(x_1,\ldots,x_{n-1},x_n)\rightarrow(x_1,\ldots,x_{n-1}+s\,x_n,x_n).$$
\end{lem}

For a bivariate polynomial system, we can compute an $s$ satisfying (\ref{eq-s}) to derive a new system $\varphi_{s,2}(f,g)$ in a generic position.

Except for computing all the complex roots of the system to get a guaranteed generic position, there is another method to check whether a sheared bivariate system $\Sigma_s=\{f(x+s\,y,y),g(x+s\,y,y)\}$ is in a generic position or not (\cite{eq2-dio}).
Let
\begin{equation} \label{eq-cs} \bar{R}_s(x)=\res_y(f(x+s\,y,y),g(x+s\,y,y)).\end{equation}
Denote its square free part as $R_s(x)$. The discriminant of $R_s(x)$ with respect to $x$ is denoted as $W(s)$. If $0\neq s_0\not\in V_\R(W)$, then $\Sigma_{s_0}$ is in a generic position.

We can modify the method as below.
\begin{lem}\label{lem-s} Use the notations as before. $\Sigma_{s_0}$ is in a generic position if $R_{s_0}(x)$ (the content is assumed to be 1) is squarefree.
\end{lem}
\noindent{\bf Proof.} It is clear that $\gcd(R_{s_0}(x),\frac{\partial R_{s_0}(x)}{\partial x})=1$ if $R_{s_0}(x)$ is squarefree, which means $s_0$ is not a zero of the discriminant of $R_s(x)$ w.r.t. $x$. So the lemma is proved. $\Box$

The following corollary is obvious from Lemma \ref{lem-s}.
\begin{cor} \label{cor-gp} A zero-dimensional polynomial system $\{f,g\}\subset \Z[x,y]$ is in a generic position if $\res_y(f,g)$ is squarefree and the leading coefficients of $f$ and $g$ w.r.t. $y$ have no common factors.
\end{cor}

It is a special case for a bivariate system. The roots of the system are simple and do not overlap when projected to $x$-axis.

\section{Bivariate Systems Solving}
In this section, we will consider a zero-dimensional bivariate polynomial system, say $\{f,g\}\subset\Z[x,y]$. If it is not zero-dimensional, $\gcd(f,g)$ is not a constant and $\res_y(f,g)=0$.

The following lemma is deduced from (\cite{ful84}).

\begin{lem}[Section 1.6 of (\cite{ful84})]\label{th-i1}
Let $f,g\in\Z[x,y]$ be in a generic position w.r.t. $y$ and
$\gcd(f,g)=1$. Denote $R_1=\res_y(f,g)$, then $\pi_1$ is a
one-to-one and multiplicity-preserving map from $\{f,g\}$ to $R_1$.
\end{lem}

One can find the definition of multiplicity in $\S$ 2, Chapter 4 in (\cite{cox}). The lemma tells us that a zero $(x_0,y_0)$ of $\{f,g\}$ has the same multiplicity as $x_0$ in $R_1=0$.  We can directly derive the corollary below from Lemma \ref{th-i1}.
\begin{cor}\label{thm-bi}
Let $\Sigma=\{f,g\}\subset\Z[x,y]$ be zero-dimensional. If we compute $s$ as (\ref{eq-s}), $\Sigma'=\varphi_{s,2}(\Sigma)=\{f(x+s\,y,y),g(x+s\,y,y)\}$ are probability 1 in a generic position w.r.t $y$. If $\Sigma'$ is in a generic position, the real root(s) of $\pi_1(\Sigma')$ in $J-s\,K$ exactly corresponds to the real root(s) of $f=g=0$ in any real root candidate $J\times K$ of $f=g=0$ including the multiplicities.
\end{cor}
\noindent{\bf Proof.} A random shearing will put the system in a generic position w.r.t. $y$ with probability 1. So the first part of the corollary is correct. The second part of the lemma is guaranteed by Lemma \ref{th-i1}. $\Box$

Even when we compute $s$ as (\ref{eq-s}), the sheared system may not
be in a generic position as we mentioned in last section. Denote
$R_2=\res_y(\varphi_{s,2}(f),\varphi_{s,2}(g))$. When two conjugate
complex roots are common roots of
$\varphi_{s,2}(f)=\varphi_{s,2}(g)=0$ on a fiber $x=\alpha$
$(R_2(\alpha)=0$ and $\alpha\in\R$), and $\alpha\in
\pi_1(\varphi_{s,2}(J\times K))$ for some real root candidate
$J\times K$ of $f=g=0$, $J\times K$ will be regarded to be containing
a real root even if it does not. Then there may be an error since we
consider only the real roots of the system and ensure only that all the real roots (not all complex roots) of
$\varphi_{s,\,2}(\Sigma)$ are in ``a generic position" (not overlap when projected to $x$-axis). But we can use
Lemma \ref{lem-s} to ensure that the systems
$\varphi_{s,\,2}(\Sigma)$ is in a generic position (for all the roots with real $x$-coordinates). It is similar
for the multivariate case.

Let $$R_2(x)=\prod\limits_{i=1}^{m}r_{i}(x)^{i},$$ where, $r_i(x)$
is the factor of $R_2(x)$ with power $i$ and $m$ is the highest
power of the factors in $R_2(x)$. By Corollary \ref{thm-bi}, the
corresponding real roots of $\V_\R(\Sigma)$ to the real roots
$0=r_i(x)$ have multiplicity $i$ if the system is in a generic
position.

Now we will show how to identify the roots in $J-s\,K$. The case
when there is no root or one root of $R_2(x)=0$ in $J-s\,K$ is
simple.  We will show how to deal with the case that two or more
real roots are inside $J-s\, K$. That means
%If on a fiber $x=\alpha$, there exist a root candidate $J\times K$ which does not satisfy the condition in Lemma \ref{lem-gp}, we need to compute a good $s$ to compute another resultant $R_2(x)$. And
there exist two or more real roots of $f=g=0$ in $J\times K$. We
need to construct the corresponding isolating boxes for them. Assume
that there are two real roots of $R_2(x)=0$ in $J-s\,K$ and
$I_i=[a_i,b_i] (b_1<a_2, i=1,2)$ are their isolating intervals. The
case for more than two real roots is similar. Assume that the
corresponding isolating boxes of the roots of $f=g=0$ are $J\times
K_i, i=1,2$. Since we know $J-s\,K_i=I_i$, $K_i=-(I_i-J)/s$. Let
$J=[c,d]$. We need to ensure that $K_i (i=1,2)$ are disjoint. It is
not difficult to find that the condition is satisfied if
$a_2-b_1>d-c$ (One can find the proof from (\cite{chenglgp})). We can
refine $J$ if needed to get the isolating boxes for the roots inside
$J\times K$.

We can also have an algebraic representation for the roots of the system: linear univariate representation. The representation has a little difference with the original representation as in (\cite{chenglur}). Let $\mathcal{I}=\{I_i\times J_i, i=1,\ldots,m\}$ be the set of all the isolating boxes of the roots of the system. If we have computed an $s$ (there exists the case that $s$ is not necessary), the linear univariate representation of a bivariate system is
$$\{\mathcal{I},(\alpha,\frac{\beta-\alpha}{s}), R_1,R_2\in \Z[x]\,| R_1(\alpha)=R_2(\beta)=0\}.$$

Based on the analysis above, we have the following algorithm for
isolating real roots of a zero-dimensional bivariate system.
\begin{alg}\label{alg-bi}
Isolate the real roots of a zero-dimensional bivariate polynomial
system. Input: $f,g\in \Z[x,y]$ such that $\gcd(f,g)=1$. Output: the
isolating intervals of the real roots of $f=g=0$ as well as the
multiplicities of the corresponding roots.
\end{alg}
\begin{enumerate}
\item Compute $R_1:=\res_y(f,g)$.
\item Isolate the real roots of $R_1=0$.
\item For each real root isolating interval $I$ of $R_1=0$, compute real root candidates of $\{f,g\}$ with Algorithm \ref{alg-candi}.
\item For all real root candidates of $\{f,g\}$, compute $s$ as (\ref{eq-s}).
\item Compute $R_2:=\res_y(f(x+s\,y,y),g(x+s\,y,y))=\Pi_{i=1}^m r_i(x)^i$.
\item For each real root candidate $J\times K$ of $\{f,g\}$, the real root(s) of $R_2=0$ in the interval $J-s\,K$ correspond(s) to
the real root(s) of $f=g=0$ in $J\times K$. Separate $J\times K$ into several isolating boxes if it contains several roots.
\item Return all the isolating boxes with the multiplicity of the corresponding roots of the system.
\end{enumerate}
\noindent{\bf Remarks for the algorithm:}
\begin{enumerate} \item
The termination of the algorithm is clear. The correctness of
the algorithm is guaranteed by Corollary \ref{thm-bi} with
probability 1.
\item
We can choose an $s_0$ such that Lemma \ref{lem-s} holds after Step
4 and set $R_2(x)=\bar{R}_{s_0}(x)$ to replace Step 5. Then the
revised algorithm is certified.
\item
Let $T(x)=\gcd(R_1,R_2)$. Then on the fiber
$x=\alpha\in\V_\R(T(x))$, the system has real root $(\alpha,0)$. We
can easily find this from the linear coordinate transformation since
$\alpha+s\,0=\alpha$.
\item
For some system $\Sigma=\{f,g\}$, if it is in a generic position and
satisfies certain conditions, we can identify its real roots without
shearing the system. Thus we can stops at Step 3. The following
lemma shows the result.
\end{enumerate}
\begin{lem}\label{lem-gp}
Let $\alpha\in V_\R(\res_y(f,g))$ and $J$ the isolating interval of
$\alpha$. If there is only one root candidate $J\times K$ and,
$f(J,y)$ or $g(J,y)$ is monotonous in $K$, then $\{f,g\}$ has at
most one real root on the fiber $x=\alpha$.
\end{lem}
\noindent{\bf Proof.} It is clear that the possible common real
roots of $f=g=0$ on the fiber $x=\alpha$ appear in $J\times K$. From
Lemma \ref{lem-mono}, $f(\alpha,y)=0$ or $g(\alpha,y)=0$ has and
only has one real root in $K$. Thus $f=g=0$ has at most one real
root in $J\times K$. \qed

This lemma can be used for speedup our real root isolation without
the second resultant computation. If on each fiber $x=\alpha$ (where
$\alpha\in V_\R(\res_y(f,g))$), the condition in Lemma \ref{lem-gp}
holds, $\{f,g\}$ can be regarded as in a generic position. And
$J\times K$ is regarded as an isolating box of a real root of
$f=g=0$. Note that there may exist the case that two conjugate
complex roots have the same real $x$ coordinate $\alpha$ and the
real root candidate of $f=g=0$ on the fiber $x=\alpha$ has no real
root(s). Then there is an error. But this case seldom happens. One special case is guaranteed by Corollary \ref{cor-gp}.

\begin{exmp}
Isolate the real roots of the system $\Sigma=\{f,g\}$, where
$f=x^2+y^2-2, g=(x-2\,y^2)^2-2$. Following Algorithm \ref{alg-bi},
we have
\begin{enumerate}
\item Compute the resultant of $f,g$, we have\\ $R_1(x)=(4\,x^2+4\,x-7)^2\,(x^2-2)^2$.
\item Isolate the real roots of $R_1=0$ with precision $2^{-10}$, we have\\ $II=[[-{\frac {1961}{1024}},-{\frac {245}{128}}],[-{\frac {1449}{1024}},-{
\frac {181}{128}}],[{\frac {117}{128}},{\frac {937}{1024}}],[{\frac{181}{128}},{\frac {1449}{1024}}]]$.
\item For each $I\in II$, compute the real root candidates of $f=g=0$, we have the candidates below.\\
$[[[-{\frac {1449}{1024}},-{\frac {181}{128}}],[-{\frac {72905}{8388608
}},{\frac {72905}{8388608}}]],[[{\frac {117}{128}},{\frac {937}{1024}}
],[-{\frac {70721}{65536}},-{\frac {141401}{131072}}]]$,\\
$[[{\frac {117}{
128}},{\frac {937}{1024}}],[{\frac {141401}{131072}},{\frac {70721}{
65536}}]],[[{\frac {181}{128}},{\frac {1449}{1024}}],[-{\frac {42621}{
2097152}},{\frac {42621}{2097152}}]]]
$.
\item Compute $S$ as (\ref{eq-si}), we have\\
{\small $S=[[-\infty ,-{\frac {23724032}{243389}}],[-{\frac {19546112}{8976759}},-{\frac {19529728}{9125193}}],[-{\frac {1050624}{2219795}},-{\frac {
1046528}{2305693}}],[-{\frac {64}{141401}},{\frac {64}{141401}}]$,\\
$[{
\frac {1046528}{2305693}},{\frac {1050624}{2219795}}],[{\frac {
19529728}{9125193}},{\frac {19546112}{8976759}}],[{\frac {23724032}{
243389}},\infty]]$.}\\
\item There are two choices for this step. One is a probability 1 algorithm and the other is a certified one.
\begin{itemize}
\item From $S$ in last step, we can choose $s=1$. Then we compute
$R_2=4\,(4\,x^4+8\,x^3-8\,x^2-44\,x-7)\,(x^2-2)^2$. And we can
denote the square-free part of $R_2$ as $\bar{R}_2$.
\item Or we compute $\bar{R}_s(x)=\res_y(f(x+sy,y),g(x+sy,y))=- ( {x}^{2}-2 ) ^{2}$ $( -16\,{x}^{4}-32\,{x}^{3}$ $+40\,{x
}^{2}-8\,{x}^{2}{s}^{2}+120\,x{s}^{2}+56\,x-49+46\,{s}^{2}+31\,{s}^{4}
 )
$,
and its square free part is $\tilde{R}_s(x)=({x}^{2}-2)(
-16\,{x}^{4}-32\,{x}^{3}+40\,{x}^{2}-8\,{x}^{2}{s}^{2}+120\,x{s}^{2}+56\,x+31\,{s}^{4}+46\,{s}^{2}-49
)$. Let $s=1$, we have
$\tilde{R}_1(x)=({x}^{2}-2)(-4\,{x}^{4}-8\,{x}^{3}+8\,{x}^{2}+44\,x+7)
$ (removing the content $4$).  It is squarefree, so we know the system $\Sigma_1=\{f(x+y,y),\,g(x+y,y)\}$ is in a
generic position from Lemma \ref{lem-s}. This guarantees that the
final result is certified.
\end{itemize}

%We compute $\bar{R}_s(x)=- \left( {x}^{2}-2 \right) ^{2} ( -16\,{x}^{4}-32\,{x}^{3}+40\,{x
%}^{2}-8\,{x}^{2}{s}^{2}$ $+120\,x{s}^{2}+56\,x+31\,{s}^{4}+46\,{s}^{2}-49
%)$. And its squarefree part is denoted as $R_s(x)$. From $S$ in last step, we can choose $s=1$. We can find that the greatest common divisor of $R_s(x)$ and $\frac{\partial R_s(x)}{\partial x}$ is a constant when $s=1$. Thus the condition in Lemma \ref{lem-s} is satisfied.
% Then we compute $R_2=\bar{R}_s(x)|_{s=1}=4\,(4\,x^4+8\,x^3-8\,x^2-44\,x-7)\,(x^2-2)^2$. And we can denote the squarefree part of $R_2$ as $\bar{R_2}$.
%

\item Since $L=[-{\frac {1449}{1024}},-{\frac {181}{128}}]-[-{\frac {72905}{8388608
}},{\frac {72905}{8388608}}]=[-{\frac {11943113}{8388608}},-{\frac
{11789111}{8388608}}]$. We can find that $\bar{R_2}$ has different
signs at the endpoints of the interval and $\frac{\partial
\bar{R_2}}{\partial x}=0$ has no roots in $L$. So $[[-{\frac
{1449}{1024}},-{\frac {181}{128}}],[-{\frac {72905}{8388608
}},{\frac {72905}{8388608}}]]$ is an isolating interval of $f=g=0$.
We can also find that the root in $L$ corresponding to $x^2-2$. So
its multiplicity is 2. And since $(x^2-2)|\gcd(R_1(x),R_2(x))$,
$(\pm \sqrt{2},0)$ are real roots of the original system from Remark
4 of Algorithm \ref{alg-bi}. Thus $[[-{\frac
{1449}{1024}},-{\frac {181}{128}}],[0,0]]$ is an isolating interval
of $f=g=0$. The other isolating intervals can be identified
similarly. Denote all the isolating intervals as $K$. {\tiny
$$K=[[[-{\frac {1449}{1024}},-{\frac {181}{128}}],[0,0]],[[{\frac
{117}{128}},{\frac {937}{1024}} ],[-{\frac {70721}{65536}},-{\frac
{141401}{131072}}]],[[{\frac {117}{ 128}},{\frac
{937}{1024}}],[{\frac {141401}{131072}},{\frac {70721}{
65536}}]],[[{\frac {181}{128}},{\frac {1449}{1024}}],[0,0]]].
$$}
Then we can get the LUR of the system:
$$\{K, (\alpha,\beta-\alpha), R_1(x),R_2(x)| R_1(\alpha)=0, R_2(\beta)=0\}.$$

\end{enumerate}
\end{exmp}
Now we consider a bivariate zero-dimensional systems with $m (>2)$
polynomials, we just take $m=3$ for an illustration, it is similar
for the case of $m>3$. Let $\Sigma=\{f,g,h\},$ where $f,\,g,\,h\in
\Z[x,\,y]$. Let
$p=\mbox{gcd}(f,g),\,f^*=\frac{f}{p},\,g^*=\frac{g}{p}$. We
have $\V_{\R}(f,g,h)=\V_{\R}(f^*,g^*,h)\cup\V_{\R}(p,h)$.
Furthermore, let $q=\mbox{gcd}(g^*,h)$ and
$g^{**}=\frac{g^{*}}{q},\,h^*=\frac{h}{q}$, then we obtain
$\V_{\R}(f^*,g^*,h)=\V_{\R}(f^*,g^{**},h^*)\cup\V_{\R}(f^*,q).$
Hence, we have
\begin{equation}\label{decomposition}
\V_{\R}(f,g,h)=\V_{\R}(f^*,g^{**},h^*)\cup\V_{\R}(f^*,q)\cup\V_{\R}(p,h)
\end{equation}
On the right side of (\ref{decomposition}), both $\{f^*,q\}$ and
$\{p,h\}$ are zero-dimensional, thus we can solve these two systems
using Algorithm \ref{alg-bi}. Now we will show how to solve the
system $\{f^*,g^{**},h^*\}$. Actually,
$\V_{\R}(f^*,g^{**},h^*)=\V_{\R}(f^*,g^{**})\cap\V_{\R}(g^{**},h^*)$.
and $\{f^*,g^{**}\}$, $\{g^{**},h^*\}$ are zero-dimensional
polynomial systems. Assume that the LUR of the systems
$\{f^*,g^{**}\}$, $\{g^{**},h^*\}$ are
\begin{equation}\label{system1}
    \big\{K_1,\,(\alpha,\frac{\beta-\alpha}{s_1})|R_{1,1}(\alpha)=0,\,R_{1,2}(\beta)=0\big\}
\end{equation}
and
\begin{equation}\label{system2}
    \big\{K_2,\,(\alpha,\frac{\beta-\alpha}{s_2})|R_{2,1}(\alpha)=0,\,R_{2,2}(\beta)=0\big\}
\end{equation}
respectively. What's more, if we chose the same value for $s$ in
equations \bref{system1} and \bref{system2}~(this can be easily
achieved), that is $s_1=s_2$ then we can get the LUR of the system
$\{f^*,g^{**},h^*\}$:
\begin{equation}\label{LUR-tri}
    \big\{K_1\cap K_2,\,(\alpha,\frac{\beta-\alpha}{s})|R_1(\alpha)=0,\,R_{2}(\beta)=0\big\},\,
\end{equation}
where
$s=s_1=s_2,\,R_1=\mbox{gcd}(R_{1,1},R_{2,1}),\,R_2=\mbox{gcd}(R_{2,1},R_{2,2})$ and in each isolating box of $K_1\cap K_2$ there exist only one real root of the system. We can also ensure that we take the same $s$ for $\V_{\R}(f^*,q)$ and $\V_{\R}(p,h)$ when solving them. Thus we can check their real roots are exactly the same or not as the roots of $\{f^*,g^{**},h^*\}$. In the end, we get all the solutions of the original system $\{f,g,h\}$.

\section{Multivariate Systems Solving}
In this section, we will show how to isolate the real roots of a general zero-dimensional polynomial system.

Let $\Sigma_n=\{f_1,\ldots,f_m\}\subset \Z[x_1,\ldots,x_n],\,m\geq
n$ be a zero-dimensional polynomial system. Let $f_i'=\sum_{j=1}^{m}
t_{i,j} f_j (1\le i\le n)$, where $t_{i,j}\in\Z$ and the rank of the matrix $(t_{i,j})$ is of full rank $n$, denoted as $\mbox{rank}(t_{i,j})=n$. Let
$g_i=\res_{x_n}(f_i',f'_n), i=1,\ldots,n-1$. Then
$\Sigma_{n-1}=\{g_1,\ldots,g_{n-1}\}$ is probability 1 to be a
zero-dimensional polynomial system. We call $\Sigma_{n-1}$ is an {\bf $(n-1)$-projection system} of $\Sigma_n$ if it is zero-dimensional. We denote $\Sigma_{n-1}=\prod_{n-1}(\Sigma_n)$ and $\prod_i(\Sigma_n)=\prod_{i}(\prod_{i+1}(\cdots\prod_{n-1}(\Sigma_n)\cdots))$. We want to mention that $\pi_i(\Sigma_n)\subset \prod_{i}(\Sigma_n)$.
Recursively, we can eliminate
variables to get a univariate polynomial. Assume that we know how to
derive the roots of $\Sigma_{i}=0$ since we know how to get the real
roots of a univariate polynomial equation or a zero-dimensional
bivariate polynomial system. Using the real root isolating intervals
of $\Sigma_i$, we can compute the real root candidates of
$\Sigma_{i+1}=\{p_1(x_1,\ldots,x_{i+1}),\ldots,p_{i+1}(x_1,\ldots,x_{i+1})\}=0$.
Computing $s$ as (\ref{eq-s}), we can get a new system
$\Sigma_{i+1}'=\{p_1(x_1,\ldots,x_{i-1},x_i+s\,x_{i+1},x_{i+1}),\ldots,p_{i+1}(x_1,\ldots,x_{i-1},x_i+s\,x_{i+1},x_{i+1})\}$.
Projecting it to obtain a zero-dimensional system
$\overline{\Sigma_{i}}$ in $i$-space as above, we can isolate its
real roots and check whether there exist real roots in the real root
candidates of $\Sigma_{i+1}=0$. Then we get the real root isolating
intervals of $\Sigma_{i+1}=0$. In a recursive way, we can obtain the
real root isolating intervals of $\Sigma_n=0$.

\begin{lem}Use the notations as above. \\
%\begin{equation}\label{eq-eqn}
\begin{center}$ \V( \prod_{i-1} (\Sigma_{i+1}))=\V(\prod _{i-1}(\Sigma_{i+1}'))\subset\V(\prod _{i-1}(\overline{\Sigma_{i}})). $
\end{center}
%\end{equation}
\end{lem}
\noindent{\bf Proof.}  The equality is true since the roots of $\Sigma_{i+1}=0$ and $\Sigma_{i+1}'=0$ have a one-to-one map and their corresponding roots differ only on the $i$-th coordinate. And the inclusion relationship is clear.
\qed

\begin{thm}\label{thm-gen} Use the notations as above and compute $s$ for the real root candidates of $\Sigma_{i+1}=0$ as (\ref{eq-s}). Then $\Sigma_{i+1}'$ is in a generic position w.r.t. $x_i,x_{i+1}$ in order $x_i\prec x_{i+1}$ with probability 1.
\end{thm}
\noindent{\bf Proof.} We can find that there are many extraneous roots in
$\Sigma_{i}=0$ corresponding to $\prod_i(\Sigma_{i+1})$. But the
number is finite. So are the roots in $\C^{i+1}$. It is similar for
$\overline{\Sigma_i}=0$ and $\prod_i(\Sigma_{i+1}')$. Thus there are
only finite complex points in $\C^{i+1}$ such that $\Sigma_{i+1}$ is
not in a generic position w.r.t. $x_i,x_{i+1}$ in order $x_i\prec
x_{i+1}$. So we prove the theorem. \qed

We give the following algorithm to isolate the real roots of a general zero-dimensional polynomial system.
\begin{alg}\label{alg-m} Isolate the real roots of a zero-dimensional polynomial system.\\
Input: $\Sigma_n=\{f_1,\ldots,f_m\}\subset \Z[x_1,\ldots,x_n]$.\\
Output: Isolating intervals of the real roots of $\Sigma_n=0$.
\end{alg}
\begin{enumerate}
\item Let $f_i'=\sum_{j=1}^{m} t_{i,j} f_j (1\le i\le n)$, where $t_{i,j}\in\Z$ and $\mbox{rank}(t_{i,j})=n$, and $\Sigma_{n-1}=\{g_1,\ldots,g_{n-1}\}$, where $g_i=\res_{x_n}(f_i',f_n'), i=1,\ldots,n-1$. In a similar way, we can get $\Sigma_{n-2},\ldots,\Sigma_1$.
\item For $i=1,\ldots,n-1$, do the following computation.
\begin{enumerate}
\item Isolate the real roots of $\Sigma_i=0$.
\item For each root isolating interval $I=I_1\times\cdots\times I_i$ of $\Sigma_i=0$,
compute root candidates of $\Sigma_{i+1}$ with Algorithm
\ref{alg-candi}.
\item Compute $S_i$ as (\ref{eq-si}) and choose $s_i$ as (\ref{eq-s}).
\item Assume that $\Sigma_{i+1}=\{p_1,\ldots,p_{i+1}\}$.
Let $p_k'=\sum_{j=1}^{i+1} t_{k,j}'
p_j(x_1,\ldots,x_{i-1},x_i+s_i\,x_{i+1},x_{i+1})$, where
$t_{k,j}'\in\Z$, $ k=1,\ldots,i+1$ and $ \mbox{rank}(t'_{k,j})=i+1$.
$\overline{\Sigma_{i}}=\{q_1,\ldots,q_{i}\}$, where
$q_k=\res_{x_{i+1}}(p_k',p_{i+1}'), k=1,\ldots,i$.
\item Isolate the real roots of $\overline{\Sigma_{i}}=0$.
\item
For each root candidate $I_1\times\cdots\times I_{i-1}\times
I_i\times K$ of $\Sigma_{i+1}$, it is a real root isolating interval
if $I_1\times\cdots\times I_{i-1}\times (I_i-s_i\,K)$ has non-empty
intersection with some real root isolating interval of
$\overline{\Sigma_{i}}=0$. It should be subdivided into two or more
isolating intervals if $I_1\times\cdots\times I_{i-1}\times
(I_i-s_i\,K)$ has intersection with two or more real root isolating
interval of $\overline{\Sigma_{i}}=0$ similarly as Step 6 in
Algorithm \ref{alg-bi}.
\end{enumerate}
\item Output the isolating boxes of $\Sigma_n$.
\end{enumerate}
\noindent {\bf Remarks:}
\begin{enumerate}
\item The termination of the algorithm is clear. The algorithm is probability 1 correct.  It is guaranteed by Theorem \ref{thm-gen}.

\item
Now we consider the LUR of $\Sigma_i (i> 2)$. Assume that we have
got the LUR for $\Sigma_{i}$. The univariate polynomials are
$T_1(y_1),\ldots, T_{i}(y_{i})$. The $s_j$'s are
$s_1,\ldots,s_{i-1}$. The real root isolating intervals are
$\mathbb{I}_{k}=I_{k,1}\times I_{k,2}\times\cdots\times I_{k,i}$,
$k=1,\ldots,p$. We know that $T_2(y_2)=0$ has a real root in
$I_{k,1}-s_1\,I_{k,2}$, $T_3(y_3)=0$ has a real root in
$I_{k,1}-s_1\,(I_{k,2}-s_2\,I_{k,3})$, \ldots, and $T_{i}(y_{i})=0$
has a real root in
$\overline{\mathbb{I}}_{k}=I_{k,1}+\Sigma_{j=1}^{i}(-1)^{j-1}s_1\cdots
s_{j-1}\, I_{k,j})$. Of course, we require that
$\overline{\mathbb{I}}_{k}$ are disjoint for any $k$. And the zeros
of $\Sigma_{i}$ can be represented as
$$\{\mathbb{I}_{k},(\alpha_1,\frac{\alpha_2-\alpha_1}{s_1},\ldots,\frac{\alpha_{i}-\alpha_{i-1}}{s_1\cdots s_{i-1}}), T_t(x)| T_t(\alpha_t)=0 \hbox{ for } t=1,\ldots, i.\}$$

\end{enumerate}

\begin{exmp}
Let's consider isolating the real roots of
$\Sigma=\{f,g,h\}=\{3\,x-y-5\,z-4,8\,x^2+8\,y^2+z^2-8,x^2+2\,y^2+4\,z^2-4\}$.

At first, we compute $p=\res_y(f,g)=209\,{x}^{2}+201\,{y}^{2}-184-6\,xy-24\,x+8\,y$, $q =\res_y(f,h)=61\,{x}^{2}+54\,{y}^{2}-36-24\,xy-96\,x+32\,y
$.

Isolate the real roots of the bivariate polynomial system
$\{p,q\}$ with Algorithm \ref{alg-bi}. Denote its LUR as
$$\{K,(\alpha,\frac{\beta-\alpha}{s_1}), T_1(x),T_2(x) |T_1(\alpha)=0, T_2(\beta)=0\},$$
where {\small $K=[[[-{\frac {433}{2048}},-{\frac {865}{4096}}],[-{\frac {15433}{16384}}
,-{\frac {123453}{131072}}]],[[{\frac {95}{256}},{\frac {761}{2048}}],
[{\frac {116549}{131072}},{\frac {58287}{65536}}]]]$},  $T_1=11667\,x^4$\,\,  $+185368\,x^2-24960\,x-48480\,x^3-14032$, $T_2=35001\,x^4-95104\,x+1203952\,x^2-393936\,x^3$ $-429504$. And $s_1=1$.

For each isolating box of $K$, compute the real root candidates of $f=g=h=0$. They are
{\tiny $$T=[[[-{\frac {433}{2048}},-{\frac {865}{4096}}],[-{\frac {15433}{16384}}
,-{\frac {123453}{131072}}],[-{\frac {96789}{131072}},-{\frac {96779}{
131072}}]],[[{\frac {95}{256}},{\frac {761}{2048}}],[{\frac {116549}{
131072}},{\frac {58287}{65536}}],[-{\frac {49489}{65536}},-{\frac {
98955}{131072}}]]].$$} We can compute a number: $s_2=1$.

The new system is $\{f',g',h'\}=\{3\,x-y-6\,z-4,8\,{x}^{2}+8\,{y}^{2}+16\,zy+9\,{z}^{2}-8,{x}^{2}+2\,{y
}^{2}+4\,zy+6\,{z}^{2}-4\}$. The resultants of $f'$ and $g',h'$ w.r.t. $z$ are $\{p',q'\}=\{90\,{x}^{2}+54\,{y}^{2}-48+36\,xy-48\,y-144\,x,369\,{x}^{2}+201\,{y}^
{2}-144+234\,xy-312\,y-216\,x
\}$. Its isolating intervals are
$K'=[[[-{\frac {433}{2048}},-{\frac {865}{4096}}],[-{\frac {106759}{524288
}},-{\frac {106637}{524288}}]]$, $[[{\frac {95}{256}},{\frac {761}{2048}}
],[{\frac {53871}{32768}},{\frac {215573}{131072}}]]]
$.

We can check that $(T[1][2]-T[1][3])\cap K'[2]\neq\emptyset$. Thus, $T[1]$ is an isolating box of the original system. Similarly, we can find that $T$ are the isolating boxes of the original system.

\end{exmp}

\section{The algorithm complexity}
In this section, we will analyze the complexity of Algorithms \ref{alg-bi}.

At first, we will introduce some notations. In what follows
$\mathcal {O}_B$ means bit complexity and the $\tilde{\mathcal
{O}}_B$-notation means that we ignore logarithmic factors. For a
polynomial $f \in \Z[X]$, $\deg (f)$ denotes its degree. By
$\mathcal {L}(f)$ we denote an upper bound on the bitsize of the
coefficients of $f$ (including a bit for the sign), sometimes we
also take the conventions in (\cite{Kerber}) that an integer
polynomial is called of magnitude $(n,\,\tau)$ if its total degree
is bounded by $n$, and each integer coefficient is bounded by
$2^{\tau}$ in its absolute value. $\O$ indicates that we omit
logarithmic factors. For $a \in \mathbb{Q}$, $\mathcal {L}(a)$ is
the maximal bitsize of the numerator and the denominator.

\begin{lem}\label{lem-mic}
Let $f\in\Z[x]$ such that $\deg(f)\le d, \mathcal{L}(f)\le \tau$. We can isolate the real
roots of $f$ using no more than $\OB(d^3\tau)$ bit operations
(\cite{pan0,michael,sch}) or $\OB(d^2\tau)$ bit
operations (\cite{pan}). We can refine all the isolating intervals to a width $2^{-L}$ or less using  $\OB(d^3\tau+d^2\,L)$ (\cite{michael}) or $\OB(d^2\tau+d\,L)$(\cite{pan1}) bit operations.%(operation complexity??not bit complexity)
%The endpoints of (all) the isolating intervals have bitsize $\mathcal{O}(d\tau)$.
\end{lem}
In this paper, we use $\OB(d^3\tau)$ for real root isolation and $\OB(d^3\tau+d^2\,L)$ for refinement of isolating intervals.

%\begin{defn}
%We denote the bit size of a real number $\xi$, which is represented
%by an interval $[a,b]$, to be $\LL(\xi)=\max{(\LL(a),\LL(b))}$.
%\end{define}

\begin{lem}[\cite{Kerber}]
Let $f(x)\in\Z[x]$ be a polynomial of $\deg_{x}(f)\leq d$,
$\LL(f)\leq \tau$, and a rational value $\frac{c}{d}$ such that $c$
and $d$ have a bitsize of at most $\sigma$, then evaluating
$f(\frac{c}{d})$ has a complexity of $\O(d(\tau+d\sigma)).$
\end{lem}
According to the lemma above, we have the following lemma directly:
\begin{lem}\label{bi-evaluation}(Rational Evaluation for bivariate
polynomials) Let $f(x,\,y)\in\Z[x,\,y]$ such that
$\deg(f)\leq d$, $\LL(f)\leq \tau$, and $\frac{c}{d}$ a rational value
 such that $\LL(c), \LL(d)\le \sigma$. Then evaluating $f(\frac{c}{d},\,y)$ has a complexity of
$\O(d^2(\tau+d\sigma))$. Moreover, $\deg(f(\frac{c}{d},\,y))\le d, \LL(f(\frac{c}{d},\,y))\le \mathcal
{O}(d\sigma+\tau)$.
\end{lem}

\begin{lem}[\cite{Kerber}]\label{sqrfree}
(Square-free part)~Let $g \in \Z[x]$ such that $\deg(g)\le d$
and $\LL(g)\le\lambda$. Its square-free part $g^{*}$ can be computed
in $\O(d^2\lambda)$. Furthermore, $\deg(g^*)\le d, \LL(g^*)\le \O(d+\lambda)$.
\end{lem}

One can find the following result in some references, such as (\citet{Kerber, rei}).
\begin{lem}\label{gcd-compute}
Let $f,g\in \Z[x]$ such that $\deg(h)\le d,\LL(h)\le\tau$ for $h=f,g$. Computing their
gcd, denoted as $p$, has a complexity of $\O(d^2\tau)$  and $\deg(p)\le d, \LL(p)\le\mathcal {O}(d+\tau)$.
\end{lem}

\begin{lem}
[\cite{basu,mignotte,yap:fund}]\label{lem-sep}Let $f(x)\in\Z[x]$ such that $\deg(f)\leq d$, $\LL(f)\leq \tau$. Then the separation bound
of $f$ is $$\sep(f)\geq
d^{-\frac{d+2}{2}}(d+1)^{\frac{1-d}{2}}2^{\tau (1-d)},$$ thus
$\log(\sep(f))=\O(d\tau)$. The latter provides a bound on the bit
size of the endpoints of the isolating intervals.
\end{lem}

\begin{lem}
[\cite{eq2-dio}]\label{lem-Dim2}Let $f, g \in (Z[y_1,\ldots , y_k])[x]$ with
$\deg_x(f)=p\geq q=\deg_x(g)$, $\deg_{y_i}(f)\leq p$ and
$\deg_{y_i}(g) \leq q$, $\mathcal {L}(f)=\tau\geq\sigma=\mathcal
{L}(g)$. We can compute $ \res_x(f, g)$ in $\tilde{\mathcal
{O}}_B(q(p+q)^{k+1}p^k\tau)$. And $\deg_{y_i}(\res_x(f,g))\leq 2pq$,
and the bit size of resultant is $\O(p\sigma+q\tau).$
\end{lem}

The following lemma shows how to compute the non-generic interval set of two isolating boxes. It will be used to bound the bitsize of $s$.

\begin{lem} \label{lem-sl} Let $L_i=J_i\times K_i=[a_i,b_i]\times[c_i,d_i]\in \IQ^2, i=1,2$ be two real root candidates of $\Sigma=0$. The widthes of $J_i, K_i$ are bounded such that  $|J_i|, |K_i|\le 2^{-D^3\tau-D^3}$, where $D (>1),\tau$ are the degree bound and the bitsize bound of the coefficients of the polynomials. Assume that $a_1\le a_2$, $c_1\le c_2$,  $a_2-b_1>2^{-D^3\tau-1}$ if $a_1\neq a_2$, $b_1=b_2$ if $a_1=a_2$, and $c_2-d_1>2^{-D^3\tau-1}$ if $c_1\neq c_2$, $d_1=d_2$ if $c_1=c_2$. Denote the non-generic interval set of $J_1\times K_1$ and $J_2\times K_2$ as $L$. Then either $L$ contains at most one integer or the integers inside $L$ is larger than  $D^6$ (less than $-D^6$).
\end{lem}
\noindent{\bf Proof.} There are three cases for the position relationship of $L_1,L_2$. We will discuss them one by one.

The first case is $a_1<a_2$ and $c_1<c_2$. When we choose $s>0$, $\varphi_{s,2}(L_1)\cap \varphi_{s,2}(L_2)=\emptyset$ is always true. So the non-generic interval set contains only negative $s$. From Formula (\ref{eqn-pro}), we have the following inequations.
$$b_1+s\,c_1\ge a_2+s\,d_2, a_1+s\,d_1\le b_2+s\,c_2.$$
Solving them we have $$\frac{a_1-b_2}{c_2-d_1}\le s\le \frac{b_1-a_2}{d_2-c_1}.$$
Since $\frac{b_1-a_2}{d_2-c_1}<\frac{b_1-a_2}{c_2-d_1}\le \frac{a_1-b_2+2^{-D^3\tau-D^3+1}}{c_2-d_1}< \frac{a_1-b_2}{c_2-d_1}+\frac{2^{-D^3\tau-D^3+1}}{2^{-D^3\tau-1}}=\frac{a_1-b_2}{c_2-d_1}+2^{-D^3+2}$, the non-generic interval set of $L_1,L_2$ contains at most one integer.

The second case is $a_1<a_2$ and $c_1=c_2$. There are two non-generic intervals for $L_1,L_2$. We consider only $s>0$ (It is similar for $s<0$). From Formula (\ref{eqn-pro}), we have
$$b_1+s\,d_1\ge a_2+s\, c_2.$$
Thus $s>\frac{a_2-b_1}{d_1-c_2}>\frac{2^{-D^3\tau-1}}{2^{-D^3\tau-D^3}}=2^{D^3-1}> D^6$. So $s> D^6 (s<- D^6)$.

The last case is $a_1=a_2$ and $c_1<c_2$. We have the following from Formula (\ref{eqn-pro})(considering only $s\ge0$).
$$b_1+s\,d_1\ge a_2+s\,c_2.$$
Thus we have $s\le \frac{b_1-a_2}{c_2-d_1}<\frac{2^{-D^3\tau-D^3}}{2^{-D^3\tau-1}}=2^{-D^3+1}$. Considering both $s\ge0$ and $s<0$, we have $-2^{-D^3+1}<s<2^{-D^3+1}$. Thus the non-generic interval set of $L_1$ and $L_2$ contains only one integer.

The lemma is proved.
\qed

\begin{thm}\label{thm-complex2}
Let $\Sigma=\{f,g\}\subset \Z[x,y]$ be a zero-dimensional polynomial
system such that $\deg(h)\le D,\LL(h)\le\tau$ for $h=f,g$. Then
we can isolate the real roots of $\Sigma=0$ with the bit complexity
$\tilde{O}_B(N^{10})$ based on Algorithm \ref{alg-bi}, where
$N=\max\{D,\tau\}$.
\end{thm}
\noindent{\bf Proof.} Following Algorithm \ref{alg-bi}, we analyze the bit complexity of each step. For the first step, the bit complexity is $\tilde{O}_B(D^4\tau)$ by Lemma \ref{lem-Dim2}.

For Step 2, the bit complexity is $\tilde{O}_B(D^7\tau)$ by Lemma
\ref{lem-mic}.  The bitsize of the endpoints of the isolating
intervals of Step 2 is $\tilde{O}_B(D^3\tau)$ by Lemma
\ref{lem-sep}.

In Step 3, when we construct the interval polynomials from the isolating
intervals derived in Step 2, the bitsize of the coefficients is bounded by
$\tilde{O}(D^4\tau)$. Thus the bit complexity of obtaining the
real root candidates of $\Sigma=0$ is bounded by
$\tilde{O}_B(D^3*D^4\tau)*D^2=\tilde{O}_B(D^9\tau)$ by Lemma \ref{lem-mic}. The bitsize of
the $y$-coordinate of the candidates is bounded by
$D*\tilde{O}(D^4\tau)=\tilde{O}(D^5\tau)$. But it can be relaxed to $\tilde{O}(D^3\tau)$ by computing the isolating intervals of
the real roots of $\res_x(f,g)=0$ directly and identifying them. The width of the isolating intervals can be regarded as $2^{-D\tau}$ by Lemma \ref{lem-mic}.

In Step 4, the number of the isolating boxes in the real root candidate set $\mathcal{J}$ is bounded by $D^3$. The bit complexity to
compute a non-generic interval set w.r.t. two candidates is $\tilde{O}_B(D^3\tau)$ since the bitsizes of the endpoints of the candidates is $\tilde{O}(D^3\tau)$ by Step 3.
The number of the different intervals of the non-generic interval set w.r.t. two candidates are at most two. In fact, for any two candidates, there is only one interval in their non-generic interval set if there $y$-coordinates are disjoint whatever their $x$ coordinates are the same or not. Otherwise, there are two connected intervals for $s>0$ and $s<0$.  So the bit
complexity to get a non-generic interval set w.r.t. $\mathcal{J}$, that is, any two real
root candidates computing non-generic interval set and joining all non-generic interval sets together, is
$D^3*(D^3-1)/2*\tilde{O}_B(D^3\tau)=\tilde{O}_B(D^9\tau)$. The
number of the intervals in the non-generic interval set w.r.t. $\mathcal{J}$ is bounded by $\tilde{O}(D^6)$. Note that the bitsizes of the endpoints of the non-generic intervals are
also $\tilde{O}(D^3\tau)$. Thus to find a generic $s$ w.r.t. $\mathcal{J}$ is bounded by $\tilde{O}_B(D^9\tau)$ bit operators. Now let us consider how to bound the bitsize of $s$. Since $\Sigma$ has at most $D^3$ root candidates, the number of the non-generic intervals w.r.t. $\mathcal{J}$ we choose is at most $D^3*(D^3-1)/2\,*2=D^6-D^3$.
%That is, there is at least one generic $s$ w.r.t.$\mathcal{J}$ in $D^6$ integers .
We can refine the real root candidates of $\Sigma=0$ such that the conditions in Lemma \ref{lem-sl} holds.
Thus there is at least one generic $s$ w.r.t.$\mathcal{J}$ in $D^6$ integers.
Note that the bit complexity of the refinement does not increase the total complexity by Lemma \ref{lem-mic}. By the result of Lemma \ref{lem-sl}, the bitsize of $s$ is bounded by $log(D^6)$ (at most $O(D^6)$ non-generic $s$ w.r.t. $\mathcal{J}$).

So in Step 5, the complexity is $\tilde{O}_B(D^4\tau)$.

In Step 6, isolating the real roots of $R_2(x)=0$ is bounded by $\tilde{O}_B(D^7\tau)$. We deal with the case that one candidate contains more than one root. Let the separation bound of $R_2=0$ be $L$. We can refine the isolating intervals of $R_2=0$ to $L/4$ and the isolating intervals of $R_1=0$ to $L/2$, the condition to ensure that the isolating boxes of the system are disjoint. The separation bound of the roots of $R_2=0$ is $\tilde{O}(D^3\tau)$. From the results in \cite{michael}, the refinements of the isolating intervals of both $R_1=0$ and $R_2=0$ are bounded $\tilde{O}_B(D^5\tau)$.

So the total complexity of the algorithm is bounded by $\tilde{O}_B(D^9\tau)$. Thus we prove the theorem.
\qed

For the certified version of Algorithm \ref{alg-bi} in its remark,
the bit complexity of computing $\bar{R}_s(x)$ and $R_s(x)$ is
bounded by $\tilde{O}_B(D^6\tau)$ according to the Lemma
\ref{sqrfree} and Lemma \ref{lem-Dim2}. We have $\deg(R_s(x))\le D^2, \LL(R_s(x))\le D^2+2\,D\tau$. Its discriminant $W(s)$ has a degree at most $D^4$. Thus, if we chose a rational value $s$, it takes at
most $D^4$ times such that $\mbox{gcd}(R_s(x),\frac{\partial
R_s(x)}{\partial x})=1$. So the bitsize of $s_0$ in Lemma
\ref{lem-s} is $4\log D$, and the bit complexity of evaluating
$R_s(x)$ at $s=s_0$ is  bounded by $\tilde{\mathcal
{O}}((D^2)^2(D^2+2D\tau+D^2*4\log D))=\tilde{\mathcal
{O}}(D^5(D+\tau))$ according to Lemma \ref{bi-evaluation}. We also have $\deg(R_{s_0}(x))\le D^2, \LL(R_{s_0}(x))\le D^2\log D+D^2+2D\tau$ which
leads to the bit complexity of computing
$\mbox{gcd}(R_{s_0}(x),\frac{\partial R_{s_0}(x)}{\partial x})$ is
$\tilde{\mathcal {O}}((D^2)^2(D^2\log D+D^2+2D\tau))=\tilde{\mathcal
{O}}(D^5(D+\tau))$. The complexity of the left part is bounded by
$\tilde{O}_B(D^{10}+D^9\tau)$. So the complexity is also
$\tilde{O}_B(N^{10})$.

For a general zero-dimensional polynomial system
$\Sigma=\{f_1,\ldots,f_m\}\subset \Z[x_1,\ldots$, $x_n]$, it is not
difficult to find that our method is double exponential.

\section{Experiments}
In this section, we compare our algorithm with some existing
methods, especially with the efficient ones. We implement our
algorithm in Maple. For the univariate solver, we can use
(\cite{eri,r-z}). We use (\cite{r-z}) in Maple. We mainly
compare with some bivariate system solvers. We compare our algorithm
(in Maple), named LUR, with local generic position (LGP, in Maple
)(\cite{chenglgp}), Hybird method(HM, in Maple)(\cite{hong2}),
Discovery (Dis, in Maple)(\cite{dis}) and Isolate (the core is in C) in
Maple(\cite{rouillier-rur}).

We have four groups of examples. Each example $\{f,g\}$ has two random dense polynomials. We get the timings from a PC with
2Quad CPU 2.66G Hz, 3.37G memory and Windows XP operating system.
We stop the computation for each solver and each system when the computing
time is larger than 500 seconds. For each case we consider 10 examples for
all solvers and get their average computing time.

For the four groups, we mainly test the influences of the degree, the multiple roots, the sparsity and the bitsizes of the coefficients of the input polynomials to the different solvers. The results are shown in Figures 1, 2, 3, 4 respectively.
The way to form examples are shown below the figures.

%One is the systems with two random dense polynomials $\{f,g\}$. They
%usually have simple roots. The coefficients of the polynomials
%ranging from -100 to 100.  The timings are in Figure 1, where the
%horizontal line is the degree of the systems and the vertical line is the computing
%times. The symbols for different solvers are marked
%in Figure 1. For other figures below, they are the same.

\begin{figure}[ht]
\centering
\begin{minipage}{0.85\textwidth}
\centering
\includegraphics[scale=0.40]{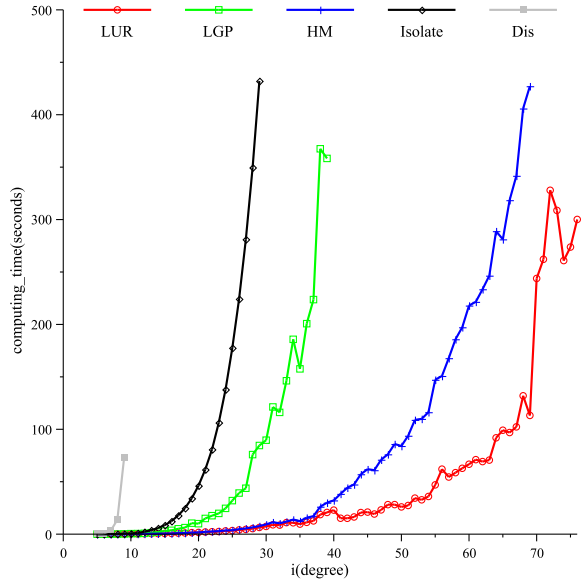}
 \caption{Timings for the system $\{f,g\}$ with simple roots, where $f(g)=randpoly([x,y],degree=i, coeffs=rand(-100..100), dense)$.
} \label{fig-simple}
\end{minipage}
\end{figure}

%One group is dense systems with multiple roots. The examples are formed as follows. Let $p,q\in\Z[x,y,z]$ be a random dense polynomial with coefficients ranging from -10 to 10. Let $f$ be the square free part of the resultant of $p,q$ w.r.t. $z$ and $g=\frac{\partial f}{\partial y}$. Usually, the system $\{f,g\}$ has multiple roots.
%The timings are in Figure 2, where the horizontal line is the sum of the degrees of $p, q$  and the vertical line is the computing times.

\begin{figure}[ht]
\centering
\begin{minipage}{0.95\textwidth}
\centering
\includegraphics[scale=0.40]{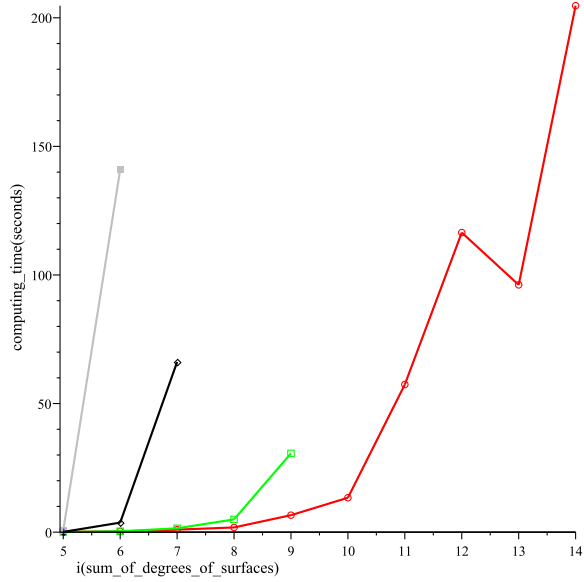}
 \caption{Timings for the system $\{f,g\}$ with multiple roots, where $p = randpoly([x,y,z],degree=ceil(i/2), coeffs=rand(-10..10), dense)$,  $q=randpoly([x,y,z],degree=i-ceil(i/2), coeffs=rand(-10..10), dense)$ and $f$ is the square free part of $\res_z(p,q)$, $g:=\frac{\partial f}{\partial y}$, where $ceil(t)$ is the minimal integer larger than a given real number $t$. The symbols for different solvers are the same as in Figure 1.
} \label{fig-slope}
\end{minipage}
\end{figure}
%We test the influence of the sparsity of the systems to the solvers.
%We choose the two polynomials with degree 15 and coefficients
%ranging from -100 to 100. The number of the terms for the
%polynomials are $5\,i$, where $i$ changes from 1 to 27. Thus the
%systems change from sparse systems to dense systems. The timings are
%in Figure 3, where the horizontal line is $i$ and the vertical line
%is the computing times.
\begin{figure}[ht]
\centering
\begin{minipage}{0.85\textwidth}
\centering
\includegraphics[scale=0.4]{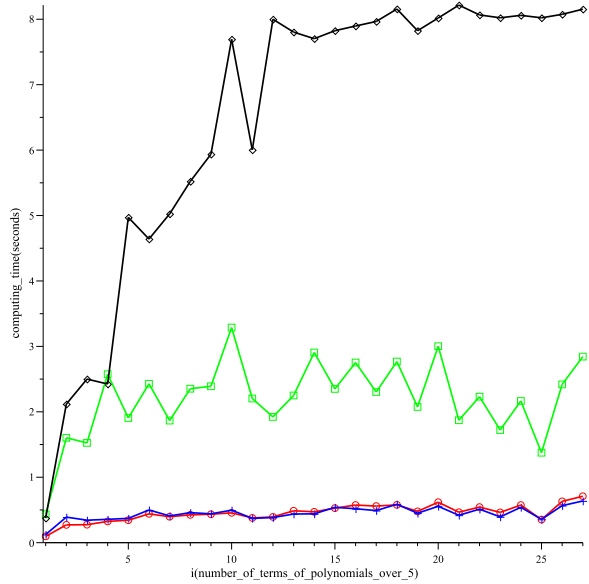}
 \caption{Timings for the system $\{f,g\}$ with simple roots, where $f(g)=randpoly([x,y],degree=15, terms=5i, coeffs=rand(-100..100), dense)$. The symbols for different solvers are the same as in Figure 1.
} \label{fig-slope}
\end{minipage}
\end{figure}

%In the last group, we study the influence of the bitsize of the
%coefficients of the systems to the solvers. The systems $\{f,g\}$ are
%formed with random dense polynomials with degree 15. Their
%coefficients range from $-2^{5\,i}$ to $2^{5\,i}$. The timings are
%in Figure 4, where the horizontal line is $i$ and the vertical line
%is the computing times. The symbols for different
%solvers are the same as in Figure 1.

\begin{figure}[ht]
\centering
\begin{minipage}{0.98\textwidth}
\centering
\includegraphics[scale=0.33]{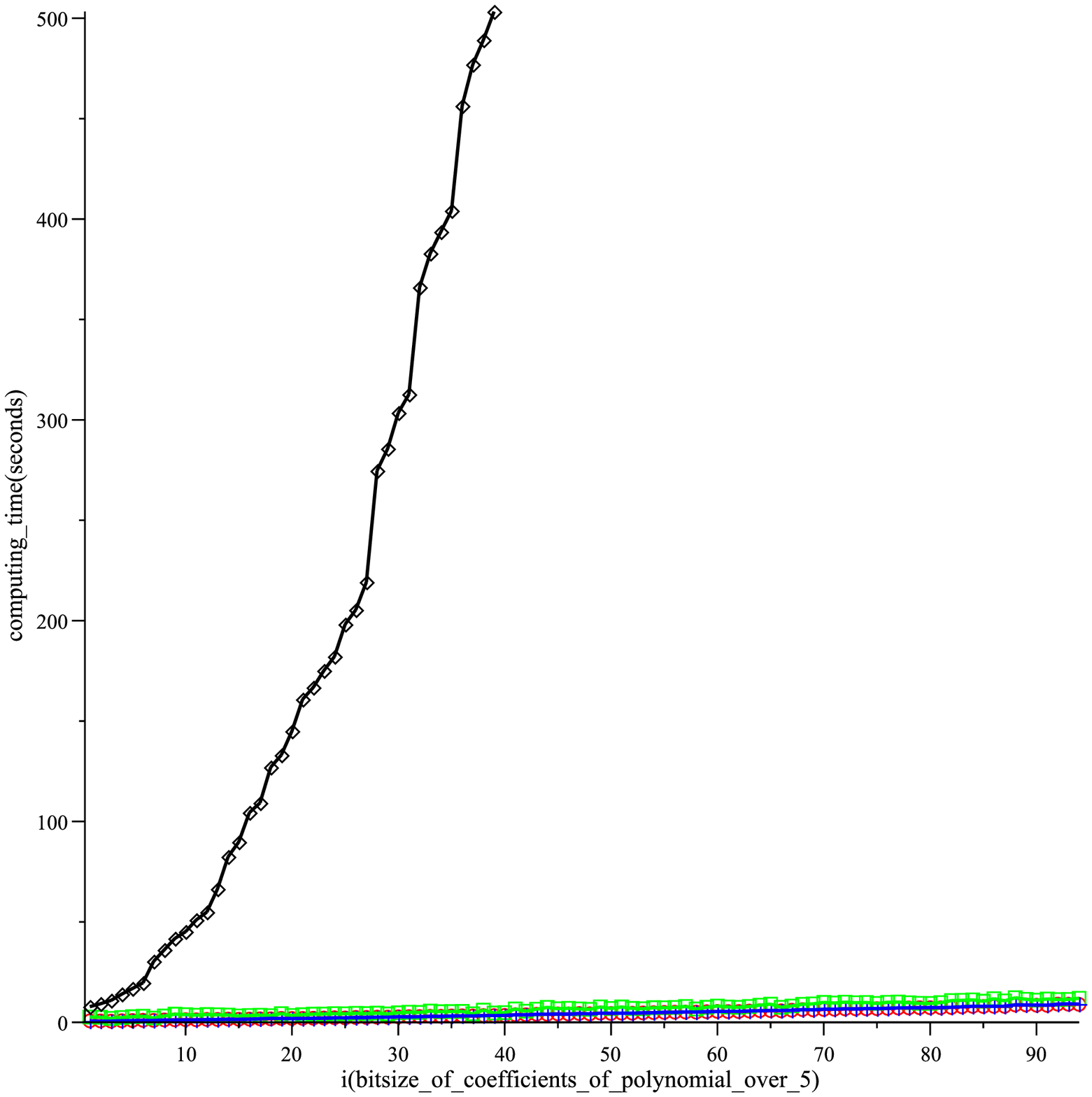}
\includegraphics[scale=0.33]{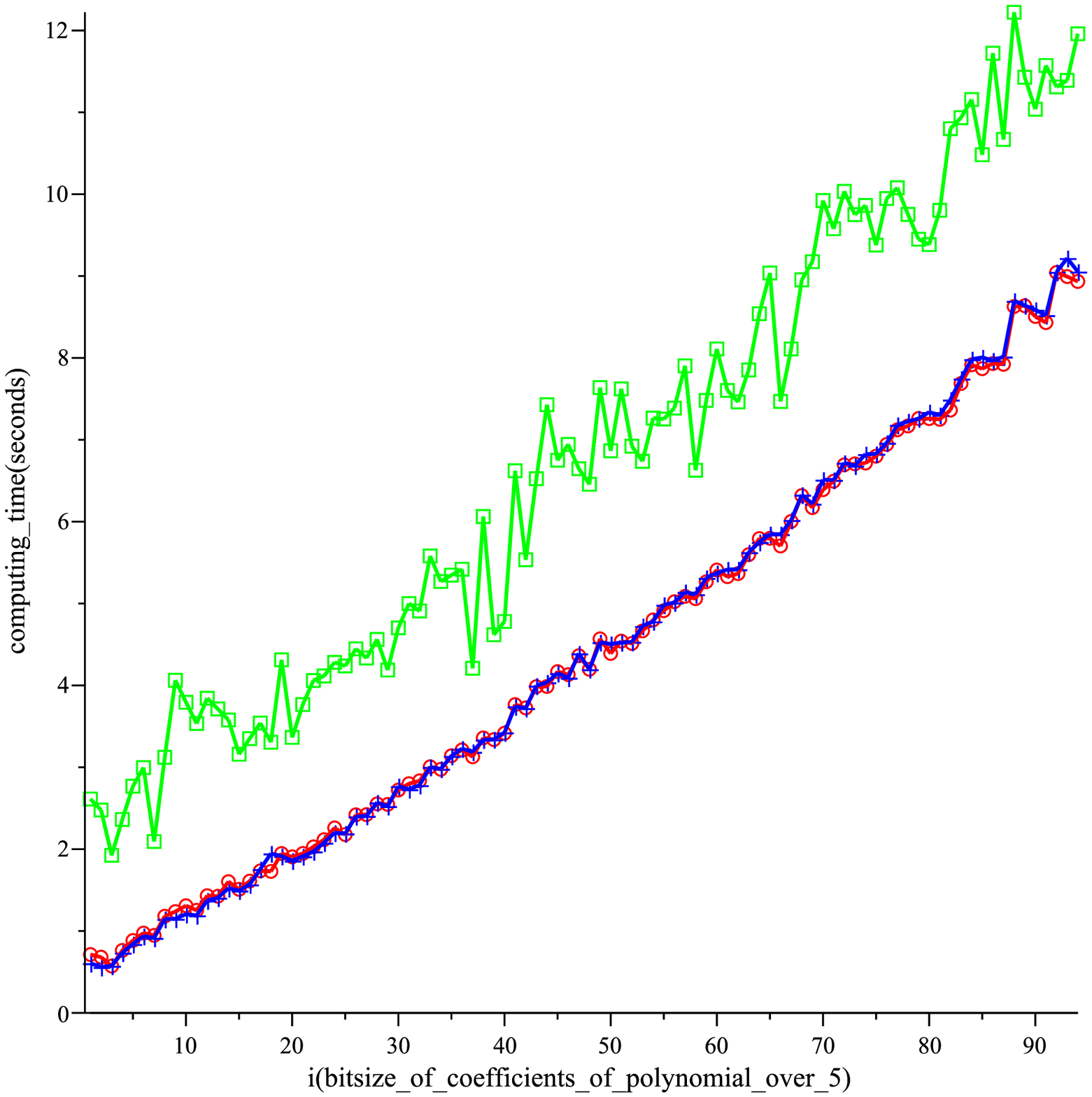}
 \caption{Timings for the system $\{f,g\}$ with simple roots, where $f(g)= randpoly([x,y], degree=15, coeffs=rand(-2^{5i}..2^{5i}), dense)$. The right one is the figure with large size without the timing for Isolate. The symbols for different solvers are the same as in Figure 1.
} \label{fig-slope}
\end{minipage}
\end{figure}

Figure 1 shows that LUR is the most efficient one among
the five solvers. Then it is HM, LGP, Isolate and Dis in decreasing order. LUR stops for a system with degree 76 because the
univariate polynomial equation solver outputs error since the
equation is out of the ability of the univariate solver.

Figure 2 shows a comparison among different solvers for systems with multiple zeros. LUR is also the most efficient one among the five solvers. It works for the systems with multiple roots of degree
$[49,48]$. Note that the bitsizes of the coefficients of the
polynomials are larger than 100. That is why LUR seems slower
comparing to itself in Figure 1. The solver HM becomes very slow for
systems with multiple roots.

From Figures 3 and 4, we can find that LUR, LGP, HM almost stable
for sparse systems. The reason is that all of them mainly involve resultant computation.
Isolate is faster for sparse systems than for dense systems.
The bitsizes of the polynomials influence all the solvers, especially for Isolate.

There is another efficient bivariate systems solver: Bisolve
(\cite{emeli1}). It is implemented in  C and use GPU parallel
technique to deal with some symbolic computations such as resultant and gcd computations. In another paper related to Bisolve, the computing times running on the same machine and the same examples were improved a lot (\cite{emeli2}) compared to (\cite{emeli1}). It is around a half computing time compared to the old one. When comparing LUR and Bisolve, we use their new data in this paper. We do not compare
with their implementation directly. But we compute the same examples
taken from (\cite{emeli2}) on our machine.
We compare the two methods in Table 1. Here, one part of  data is
taken from (\cite{emeli2}) directly. The other part of data is derived
by running on our machine. Please see Table 1
for the details. We denote our machine as M2, theirs as M1 for
convenience. We can find that LGP runs the same examples on M2
take around twice computing times (but a little less than ) as on M1
in the average level. In (\cite{emeli2}), they used some filtering
techniques to validate a majority of the candidates early. BS means
without filters, BS+all means with all filters enabled. For BS+all,
there are two groups of data. One uses GPU, denoted as
GPU-BS+all, the other does not, denoted as CPU-BS+all. BS in the table
means BS using GPU, denoted as BS+GPU. For LUR, we list the times of
computing the first resultant and isolating its real roots, denoted
as $\mathrm{T}_1$ in Table \ref{bi-variate}. The total computing
time is denoted as T.

Through we do not compare Bisolve with LUR directly, we compare them in an indirect way.
The data in their paper shows that the filtering
techniques improved the computing times a lot (usually more than one half) for Bisolve, especially for the systems with large bitsizes in coefficients. The parallel technique improved the Bisolve a lot (usually more than one half), but the improvement was not remarkable for systems with large bitsizes in coefficients. LGP is tested on both M1 and M2. The computing times of LGP on M1 are always around one half faster than on M2 for the same examples.
%The former is around one half (but a little less than) of the later.
We can find that LUR is usually faster than LGP, except for one or two examples. For some examples, LUR on M2 is faster than BS, CPU-BS+all on M1. The bitsizes of the coefficients of the systems influence BS and LUR deeper than GPU-BS+all and CPU-BS+all. We can find that for many examples, the total computing times of GPU-BS+all are less than the computing times for computing only the first resultant and its real root isolation. We use the computing times of GPU-BS+all and LGP to get a rate on M1, denoted as R1. Similarly, we can get R2 for LUR and LGP on M2. We can find that R1 is usually less than R2 except for some examples. The average level is around R1: R2 $\approx$ 1: 3. Note that the part of computing resultants, real root isolation and computing $s$ in LUR can be parallelized. Considering the influence of machines, parallel techniques and coding languages, our algorithm can be improved a lot.
%LUR should be faster than Bisolve under the same conditions.

From the comparisons before, we can conclude that LUR is efficient and stable for zero-dimensional bivariate polynomial systems.

{\tiny
\begin{table}\scriptsize
\begin{center}
\begin{tabular}{|c|c|c|c|c|c|c|c|c|c|}
  \hline
  \multicolumn{10}{|l|}{\hspace{2.5cm} comparing the computing times of Bisolve and LUR on special curves} \\
  \hline
  \multirow{3}*{ Machine} & \multicolumn{5}{l|}{\scriptsize{Linux platform on a $2.8$ GHz }}&\multicolumn{4}{l|}{\scriptsize{Win XP on Inter(R) Core}(TM) }\\

  &\multicolumn{5}{l|}{\scriptsize{$8$-Core Inter Xeon W$3530$ }}&\multicolumn{4}{l|} {\scriptsize{ 2 quad CPU Q9400 @2.66GHz }}\\

  &\multicolumn{5}{l|}{\scriptsize{with $8$MB of L$2$
  cache}}&\multicolumn{4}{l|} {\scriptsize{with 2$\times$3MB of L2 cache}}\\

  \hline
  Code language & \multicolumn{4}{|c|}{C$++$}&\multicolumn{5}{|c|}{Maple }\\
  \hline
  GPU speedup &\multicolumn{3}{c|}{YES}&\multicolumn{6}{c|}{NO}\\
  \hline
    \multirow{2}*{curves} & \,\,\multirow{2}*{\tiny BS}\,\,\, & \multirow{2}*{\tiny BS$+$all} &\multirow{2}*{\small $\frac{\mathrm{BS+all}}{ \mathrm{LGP}}$}& \multirow{2}*{\tiny BS$+$all} & \multirow{2}*{\tiny LGP} & \multirow{2}*{\small $\frac{\mathrm{LUR}}{ \mathrm{LGP}}$}& \multirow{2}*{\tiny LGP} &
 \multicolumn{2}{c|}{\tiny LUR}\\
 \cline{9-10} & & & & & & & &\,\,\,{\tiny $T_1$}\,\,\,& {\tiny $T$} \\
  \hline
  13\_sings\_9 & 2.13& 0.97&0.35& 1.65&2.81 &0.83 & 4.78& 1.78& 3.95\\
  \hline
  FTT\_5\_4\_4 & 48.03& 20.51&0.10 &52.21& 195.65&0.18 & 279.48& 2.20& 50.34\\
  \hline
  L4\_circles & 0.92& 0.74&0.10 & 1.72&7.58& 0.16& 13.86& 0.49& 2.22\\
  \hline
  L6\_circles & 3.91& 2.60&0.05 &16.16&51.60 & 0.18& 47.45&2.33 &8.77 \\
  \hline
  SA\_2\_4\_eps & 0.97& 0.44&0.09 & 4.45&4.69& 0.89& 8.92& 2.20& 7.92\\
  \hline
  SA\_4\_4\_eps & 4.77& 2.01& 0.04& 91.90&54.51& 1.15& 88.63& 12.23&102.17 \\
  \hline
  challenge\_12 & 21.54&7.35 & 0.20& 18.90& 37.07&0.85& 57.20& 4.45& 48.63\\
  \hline
  challenge\_12\_1 & 84.63& 19.17& 0.07& 72.57&277.68& 0.32& 385.28& 7.99&123.86 \\
  \hline
  compact\_surf & 12.42& 4.06& 0.34& 12.18& 12.00& 2.81&15.39& 2.20& 43.19\\
  \hline
  cov\_sol\_20 & 28.18& 5.77& 0.03& 16.57& 171.62& 0.03& 393.84&5.11&12.97 \\
  \hline
   curve24& 85.91& 8.22& 0.22& 25.36&37.94 & 0.21& 65.11&6.56& 13.75\\
\hline
   curve\_issac& 2.39& 0.88& 0.02&1.82 & 3.29&0.39&6.39 & 0.63&2.47 \\
\hline
cusps\_and\_flexes   & 1.17& 0.63& 0.26&1.27& 2.43& 0.83& 5.47&1.78 &4.56 \\
   \hline
degree\_7\_surf   & 29.92& 7.74& 0.06& 90.50& 131.25&0.14& 203.30& 10.58&28.80 \\
   \hline
dfold\_10\_6   &3.30 & 1.55& 0.41& 17.85& 3.76&0.50 & 6.19&0.13&3.08 \\
   \hline
grid\_deg\_10   &2.49 & 1.20& 0.45&2.49 & 2.64& 0.71& 6.06&2.19&4.30 \\
   \hline
huge\_cusp   & 9.64& 6.44& 0.06&13.67 &116.67 & 0.41& 224.98&76.00& 91.28\\
   \hline
mignotte\_xy   & {\tiny timeout} & 243.16&- & 310.13& {\tiny timeout}& - &{\tiny timeout} &322.00& 325.08\\
   \hline
spider   & 167.30& 46.47& - & 216.86& {\tiny timeout}& - & {\tiny timeout}& 101.19&202.02\\
   \hline
swinnerton\_dyer   & 28.39& 5.28& 0.19&24.38&27.92 & 1.10& 46.36& 1.03& 51.00\\
   \hline
ten\_circles   & 4.62& 1.33& 0.27&3.74 & 4.96& 0.54&9.09&0.55 & 4.95\\
   \hline\hline
15,\,10,\,dense& 56.40&1.55&0.27&2.66&5.65&0.29&13.49&1.84&3.89\\
\hline
15,\,128,\,dense& 95.35&2.01&0.19&2.30&10.46&0.38&21.50&5.94&8.20\\
\hline
15,\,512,\,dense& 195.01&3.95&0.12&4.22&33.87&0.46&28.27&12.30&13.06\\
\hline
15,\,2048,\,dense& {\tiny timeout}&19.89&0.10&20.45&190.86&0.45&233.13&100.58&105.58\\
\hline
15,\,10,\,sparse& 3.66&1.00&0.44&1.39&2.25&0.30&4.49&0.69&1.33\\
\hline
15,\,128,\,sparse& 12.14&1.25&0.29&1.35&4.27&0.38&8.83&2.73&3.36\\
\hline
15,\,512,\,sparse& 43.36&2.54&0.16&2.54&15.48&0.45&28.72&12.22&12.95\\
\hline
15,\,2048,\,sparse& 408.90&10.97&0.12&10.98&89.35&0.61&245.14&148.19&150.49\\
\hline
\end{tabular}
\end{center}
\caption{Timings for multivariate case: the system are formed by
random dense polynomials with the given degrees}\label{bi-variate}
\end{table}
}

\begin{table}[!hbp]
\centering
\begin{tabular}{|c|c|c|c|}
\hline
Degree Type
& LUR &  Isolate & Dis \\
\hline
 [3, 3, 3] &0.7644 &0.0874& 340.7092 \\
\hline
[5, 5, 5]& 27.1829 & 3.8826 & - \\
\hline
[2, 9, 9]& 10.1908 & 11.7686 & -  \\
\hline
[7, 7, 7]& 614.1030 & 106.7302 & - \\
\hline
[3, 15, 15]& 498.4531 & 1720.2013& - \\
\hline
\end{tabular}
\caption{Timings for multivariate case: the system are formed by
random dense polynomials with the given degrees}\label{multi-variate}
\end{table}

We also compare LUR with other efficient solvers for multivariate polynomial systems. We compare mainly with Dis and Isolate, see Table
\ref{multi-variate}. LUR is always faster than Dis. When there is a
polynomial with lower degree in the system, LUR is faster than
Isolate and it is slower than Isolate for the systems with equal
degrees. The reason is that the former case can be projected to a
bivariate system of lower degree. For the system with more
variables, it is similar. Note that the core of Isolate is in C, ours is in Maple. For the same algorithm, the implementation in C is usually several times faster than Maple.

\section*{Acknowledgement} The authors would like to thank Prof. Xiao-Shan Gao for his good advices on the paper. All the authors would like to thank the anonymous referees, their suggestions improve the paper. The work is partially supported by NKBRPC
(2011CB302400), NSFC Grants (11001258, 60821002, 91118001), SRF for ROCS, SEM, and China-France
cooperation project EXACTA (60911130369).

%\vskip 10pt
%\begin

%\bibliographystyle{abbrv}
%\bibliography{fg10bib}
%

\end{document}